\def\half{\frac{1}{2}}
\def\({\left (}
\def\){\right)}
\def\[{\left [}
\def\]{\right]}
\def\<{\left <}
\def\>{\right>}
\documentclass[a4paper,12pt]{article}
\usepackage{amssymb, amsmath}
\usepackage{epsfig}
\makeatletter
\renewcommand{\section}{{\setcounter{equation}{0}}\@startsection%
{section}%
{1}%
{0mm}%
{-\baselineskip}%
{0.5\baselineskip}%
{\normalfont\normalsize\bfseries}%
} \makeatother
\makeatletter
\renewcommand{\subsection}{\@startsection%
{subsection}%
{2}%
{0mm}%
{-\baselineskip}%
{0.5\baselineskip}%
{\normalfont\normalsize\bfseries}}%
\renewcommand{\theequation}{\arabic{section}.\arabic{equation}}

\newcommand{\ds}{\displaystyle}
\newcommand{\ben}{\begin{enumerate}}
\newcommand{\een}{\end{enumerate}}
\newcommand{\be}{\begin{equation}}
\newcommand{\ee}{\end{equation}}
\newcommand{\bes}{\begin{equation*}}
\newcommand{\ees}{\end{equation*}}
\newcommand{\bea}{\begin{eqnarray}}
\newcommand{\eea}{\end{eqnarray}}
\newcommand{\beas}{\begin{eqnarray*}}
\newcommand{\eeas}{\end{eqnarray*}}
\newcommand{\begth}{\begin{theorem}}
\newcommand{\enth}{\end{theorem}}
\newcommand{\blem}{\begin{lemma}}
\newcommand{\elem}{\end{lemma}}
\newcommand{\bpr}{\begin{proof}}
\newcommand{\epr}{\end{proof}}
\newcommand{\non}{\nonumber}
\newcommand{\nl}{\newline}
\newtheorem{remark}{Remark}[section]

\newtheorem{theorem}{Theorem}[section]
\newtheorem{lemma}{Lemma}[section]

\newenvironment{proof}{{\bf Proof:\ \ }}{\hfill$\square$\vskip.0cm}
\def\RR{\mathbb{R}}

\def\CC{\mathbb{C}}

\def\ZZ{\mathbb{Z}}

\def\la{\langle}
\def\ra{\rangle}
\def\eps{\epsilon}
\def\sLambda{{\hbox {\tiny{$\Lambda$}}}}
\usepackage{color}

\newcommand{\tit}{\textit}
\def\*{{\phantom *}}
\begin{document}
\markboth{Proof of the Variational Principle for a Pair Boson Model}
{Proof of the Variational Principle for a Pair Boson Model}
\vskip 1.5cm
\begin{center}
{\bf Proof of the Variational Principle for a Pair Hamiltonian Boson Model}
\linebreak \vskip 1cm
{\bf Joseph V. Pul\'e\,}\footnote{{\tit
Research Associate, School of Theoretical Physics, Dublin Institute
for Advanced Studies.}} \linebreak School of Mathematical Sciences
\linebreak University College Dublin\\Belfield, Dublin 4, Ireland
\linebreak Email: Joe.Pule@ucd.ie \vskip 0.1cm
and \vskip 0.1cm {\bf Valentin A. Zagrebnov} \linebreak Universit\'e de la
M\'editerran\'ee and Centre de Physique Th\'eorique \linebreak
Luminy-Case 907, 13288 Marseille, Cedex 09, France \linebreak Email:
zagrebnov@cpt.univ-mrs.fr
\end{center}
\vskip 0.5cm
\begin{abstract}
\vskip -0.7truecm
\mbox{}

\noindent We give a two parameter variational formula for the
grand-canonical pressure of the \textit{Pair Boson Hamiltonian}
model. By using the Approximating Hamiltonian Method we provide a
rigorous proof of this variational principle.
\\
{\bf  AMS :}
82B10, 
82B26, 
82B21, 
81V70  

\end{abstract}

\textbf{Contents}

\vskip 0.1cm

1. Introduction \\
2. Superstability \\
3. The First Approximation \\
4. The Second Approximation \\
5. Discussion \\
Acknowledgements \\
Appendix A: Commutators \\
Appendix B: Bounds \\
References

\newpage\setcounter{page}{1}
\section{Introduction}\label{Intr}

The first version of the \textit{Pair Boson Hamiltonian} (PBH) model was proposed
by Zubarev and Tserkovnikov in 1958 \cite{ZuTs}.
Their intention was to generalize the Bogoliubov model of the Weakly
Imperfect Bose Gas \cite{Bog1} by including more terms from the
total interaction, without losing the possibility of having an exact
solution. We refer the reader to \cite{Lieb} and to \cite{ZagBru}
for a more recent discussion of this question.
\\
The suggestion of  Zubarev and Tserkovnikov \cite{ZuTs} was to
consider a truncated Hamiltonian which includes a \textit{diagonal}
term representing forward-scattering and exchange-scattering as well
as a \textit{non-diagonal} BCS-type interaction term. The model
containing only the forward-scattering part of the interaction
corresponds to the Mean-Field (or the Imperfect) Bose gas, see
\cite{ZagBru} and \cite{PZ-Mean} for details. Using the same method
as they had used earlier for the fermion BCS model \cite{BogZuTs},
the authors give in \cite{ZuTs} a \lq\lq solution" of the PBH model.
Later this Hamiltonian became the subject of very intensive analysis
\cite{GirArn}-\cite{Lu}, leading essentially to the same conclusion
as in \cite{ZuTs}, namely, that the PBH has the same thermodynamic
properties as a certain \textit{approximating} Hamiltonian quadratic
in the creation and annihilation operators. Using this Hamiltonian
which can be diagonalized by the canonical Bogoliubov
transformation, its thermodynamic properties were investigated and
it was shown to have some intriguing properties. One of these is
possibility of the occurrence of two kinds of condensation, the
standard one-particle Bose-Einstein condensation as well as a
BCS-type pair condensation which may appear in two stages, see e.g.
\cite{Ko}, \cite{IaMarVas}. Another one concerns the gap in the
spectrum of \lq\lq elementary excitations" \cite{GirArn}-\cite{Lu}.
In spite of fairly convincing arguments these papers did not prove
rigorously that the above mentioned solution of the PBH model is
exact. A mathematical treatment of the PBH model, related to
representations of the Canonical Commutation Relations
(\textit{CCR}) appeared in \cite{EzLu}.
\par
In the present paper we give a variational formula
for the pressure for the PBH model and provide a rigorous derivation of the formula.
The latter yields the same expression for the
pressure as was obtained in \cite{ZuTs}, the
corresponding Euler-Lagrange equations coinciding with
self-consistency equations studied in \cite{ZuTs} and
\cite{GirArn}-\cite{EzLu}. In an earlier paper \cite{PZ-Pair} we
conjectured that the pressure can be expressed as the
\textit{supremum} of a variational functional depending on two
measures: a positive measure describing the particle density and a
complex measure describing the pair density, similar to the Cooper
pairs density in the BCS model. This confirmed the conclusion of
\cite{Ko}, \cite{IaMarVas} about the coexistence of one-particle and
pair condensates. The study
in \cite{PZ-Pair} was inspired by the \textit{Large Deviation
Principle} (LDP) developed for the analysis of boson systems in
\cite{BLP1}-\cite{DLP2}. This method gives rigorous results for
the pressure in the case of models with \textit{diagonal}
(commutative) boson interactions. A similar technique was developed
in \cite{CLR}-\cite{RW} based on the work \cite{PRV}, extending the LDP to \textit{noncommutative}
Mean-Field models (including the BCS one) with only bounded
operators involved in Hamiltonians. Since neither of these methods
apply to the PBH without extensive modifications, here we opted for the
\textit{Approximating Hamiltonian Method} (AHM) \cite{BBZKT}, which
has been already successfully applied to many models, including some
interacting boson models (see for example \cite{ZagBru},
\cite{PZ-Mean}, \cite{Gi}).
\par
There is renewed interest in the properties of the PBH interaction in the
context of finite boson systems confined in a
magneto-optic trap, see e.g. \cite{DuSch}-\cite{Ovch}. We do not
discuss this aspect in the framework of our approach leaving it for
future publications.
\par
Now we turn to the exact formulation of the PBH model in its
simplest form, that is, with \textit{constant} pair and mean-field boson couplings
\cite{PZ-Pair}.
\par
Let $\Lambda\subset \RR^\nu$ be a cube of volume $V
=\left|\Lambda\right|$ centered at the origin. Then the kinetic
energy operator for a particle of mass $m$ confined to the cubic box
$\Lambda$, that is the operator $-\Delta/2m$ with periodic boundary
conditions, has eigenvalues $\eps(k)=\|k\|^2/2m$, $k\in
\Lambda^*:=\{2\pi s/V^{1/\nu}| s\in\ZZ^\nu\}$. Consider a system of
identical bosons of mass $m$ enclosed in $\Lambda$. For $k\in
\Lambda^*$ let $a^*_k$ and $a^\*_k$ be the usual boson creation and
annihilation operators satisfying the \textit{CCR} $ \ [a^\*_{k{\phantom '}},a^*_{k'}]=\delta^\*_{k,k'}$
and let $N_k:=a^*_ka^\*_k $ be the $k$-mode particle number operator. The
kinetic-energy operator $T_\sLambda$ for the \textit{Perfect
Bose-gas}, can be expressed in the form $T_\sLambda:=\sum_{k\in
\Lambda^*}\eps(k)N_k$.
\\
To introduce a \textit{pairing term} in the Hamiltonian we shall
need the operators
\be \label{A-oper}
A_k =A_{-k}:=a^\*_k a^\*_{-k} \,,\,\,\,\,\,k\in \Lambda^*\,.
\ee
Let
\be \label{N-Qtilde-oper}
N_\sLambda:=\sum_{k\in \Lambda^*}N_k \,\,\,\,\, \mbox{and}
\,\,\,\,\, {\tilde Q}^\*_\sLambda:=\sum_{k\in \Lambda^*}{\tilde \lambda}(k) A_k \,,
\ee
where the function ${\tilde \lambda}:\RR^\nu\mapsto \CC$ satisfies the following conditions:
\begin{equation*}
|{\tilde \lambda}(k)|\leq |{\tilde \lambda}(0)|= 1 \,, \
{\tilde \lambda}(k)={\tilde \lambda}(-k) \ \ \mbox{for all} \ \ k\in \RR^\nu \,,
\end{equation*}
there exists $\mathfrak{C}<\infty$ and $\delta >0$ such that
\be
\label{lambda bound}
|{\tilde \lambda}(k)|\leq \frac{\mathfrak{C}}{1+\|k\|^{\max(\nu,\,\nu/2+1)+\delta}}
\ee
for all $k\in \RR^\nu$.
Note that (\ref{lambda bound}) implies that ${\tilde \lambda}\in L^1(\RR^\nu)$ and that
there exists $M<\infty$ such that
\be
\label{M}
\mathfrak{m}_\sLambda:=\sum_{k\in \Lambda^*}|{\tilde \lambda}(k)| \leq MV,
\ee
\be
\label{M2}
\mathfrak{n}_\sLambda:=\sum_{k\in \Lambda^*}\eps(k)|{\tilde \lambda}(k)|^2 \leq MV,
\ee
and
\be
\label{M3}
\mathfrak{c}_\sLambda:=\sup_{k\in \Lambda^*}\eps(k)|{\tilde \lambda}(k)|^2 \leq M
\ee
for all $\Lambda \subset \RR^\nu $.
\nl
Then for \textit{constant} couplings $u, v$ the PBH is defined by
\be
H_\sLambda:=T_\sLambda- \frac{u}{2V}\, {\tilde Q}^*_\sLambda {\tilde Q}^\*_\sLambda
+\frac{v}{2V}\, N_\sLambda^2.
\label{pre-hamiltonian}
\ee
\begin{remark}\label{lambda}
Let $\varphi:=\arg {\tilde \lambda}(0)$ and
$\lambda(k):={\tilde \lambda}(k)e^{-{\rm i}\varphi}$. Then
$\lambda(0)=1$ and we can write $H_\sLambda$ in the form
\be
H_\sLambda=T_\sLambda- \frac{u}{2V}\, Q^*_\sLambda Q^\*_\sLambda
+\frac{v}{2V}\, N_\sLambda^2
\label{hamiltonian} \ee
with
\be \label{N-Q-oper}
Q^\*_\sLambda:=\sum_{k\in \Lambda^*}\lambda(k) A_k
\,, \ee
where $|\lambda(k)|\leq \lambda(0)= 1$ for all $k\in
\RR^\nu$.
\end{remark}
\begin{remark}\label{u-v-conditions}
We shall assume that $v>0$ and $\alpha:= v - u > 0$. The latter
condition ensures the superstability of the model, see Theorem
\ref{super}. Note that in the case $u\leq 0$ {\rm{(}}BCS repulsion{\rm{)}}, the second condition $\alpha > 0$
is trivially satisfied. In \cite{PZ-Pair} we have proved that the case $u\leq
0$ gives the same thermodynamics as the Mean-Field {\rm{(}}MF{\rm{)}} Bose-gas:
\begin{equation}\label{MF-Hamil}
H^{MF}_\sLambda:=T_\sLambda + \frac{v}{2V}\, N_\sLambda^2 \,.
\end{equation}
Thus in deriving the variational formula we emphasize the case
$u>0$. We recall that this condition is necessary for nontrivial
condensation of \textit{boson pairs}, see e.g.
\cite{Wen}-\cite{PZ-Pair}. We shall discuss the relation between
these conditions and the thermodynamic properties of the model
(\ref{hamiltonian}) in Section \ref{Discussion}.
\end{remark}
For the convenience of the reader we now state (without proof) the
principal theorems and describe the the logical sequence used in
proving the main result of this paper. We shall need the
grand-canonical pressures for several \textit{approximating}
Hamiltonians. Recall that for an inverse temperature $\beta$ and
a chemical potential $\mu$ the the grand-canonical pressure for a
system with Hamiltonian $\mathcal{H}_{\sLambda}$ is
\be
\label{press}
\frac{1}{\beta V}\ln \mbox{Tr} \exp
\left\{-\beta (\mathcal{H}_{\sLambda} - \mu N_{\sLambda})\right\} \ .
\ee
For simplicity in the sequel we shall omit the thermodynamic variables
$\beta$ and $\mu$ and we shall write, for example, $p_\sLambda$ for
the grand-canonical pressure corresponding to the Hamiltonians
$H_\sLambda$
\be \label{press-main}
p_\sLambda:=\frac{1}{\beta V}\ln \mbox{Tr} \exp
\left\{-\beta (H_{\sLambda} - \mu N_{\sLambda})\right\}.
\ee
We shall denote the thermodynamic limit $\Lambda \uparrow
\mathbb{R}^\nu$ by the symbol \lq\,${\ds \lim_\Lambda}$\,'.
\par
Consider the \tit{approximating} Hamiltonian
\be \label{Approx-2-eta=0}
H^{(2)}_\sLambda(q,\rho):=T_\sLambda+v\rho
N_\sLambda-\frac{1}{2}u(Q^*_\sLambda q + Q^\*_\sLambda q^*)
-\frac{V}{2}v\rho^2 +\frac{V}{2}u |q|^2  \ ,
\ee
where $q\in\CC$ and $\rho\in \RR_+$ are variational parameters.
The Hamiltonian $H^{(2)}_\sLambda(q,\rho)$ can be diagonalized and the corresponding
pressure $p^{(2)}_\sLambda(q,\rho)$ can be calculated explicitly to give in the
thermodynamic limit
\bea
\label{lim-press-2nd Approx-eta=0}
p^{(2)}(q,\rho):&=&
\lim_{\Lambda}p^{(2)}_\sLambda(q,\rho)\non\\
&=&\int_{\RR^\nu}
\frac{d^\nu k}{(2 \pi)^\nu} \left \{-\frac{1}{\beta } \ln [1-\exp(-\beta
E(k,\!q,\!\rho))]
-\frac{1}{2}\(E(k,\!q,\!\rho) -f(k,\!\rho)\)\right\}\non \\
&& \hskip 1cm -\frac{1}{2}u |q|^2 +\frac{1}{2}v\rho^2 \ ,
\eea
where
\be \label{E}
E(k,\!q,\!\rho):= \{f^2(k,\!\rho)-|h(k,\!q)|^2 \}^{1/2}\,\ ,
\ee
with
\begin{equation}\label{fh}
f(k,\!\rho):=\eps(k)-\mu +v\rho\ \ \ \ {\rm and} \ \ \ \
h(k,\!q):= u\, q\, \lambda^{*}(k) \ .
\end{equation}
Using (\ref{Approx-2-eta=0}) the Hamiltonian
(\ref{hamiltonian}) can be written identically as
\be
H_\sLambda=H^{(2)}_\sLambda(q,\rho)+H^{r}_\sLambda(q,\rho) \ee where
\be
H^{r}_\sLambda(q,\rho):=-\frac{1}{2V}u(Q^*_\sLambda-Vq^*)(Q^\*_\sLambda-Vq)
+\frac{1}{2V}v (N_\sLambda-\rho)^2 \ .
\ee

The \textit{main result} of
this paper states that if the \textit{variational
parameters} $q$ and $\rho$ are chosen in an \lq\lq optimal" way,
then the contribution to the pressure arising from the
\textit{residual} term $H^{r}_\sLambda(q,\rho)$ vanishes in the
thermodynamic limit.
\par
Let us define the following function for $q\geq 0$ and $\rho\geq 0$
\be
\label{inf sigma}
\sigma(q,\rho):=\inf_{k\in \RR^\nu}\(f(k,\rho)-|h(k,q)|\)=v\rho -\mu -|u|q \,,
\ee
see (\ref{fh}).
\begin{theorem} \label{Main Theorem}
The limiting pressure for the {\rm PBH} model (\ref{hamiltonian})
with $u>0$ {\rm{(}}BCS attraction{\rm{)}}
has the form
\be \label{lim-Var Princ} p:=\lim_\Lambda p_\sLambda
=\sup_{q\in\CC}\inf_{\rho\geq 0}p^{(2)}(q,\rho)=
\sup_{q\geq 0}\inf_{\rho\,:\,\sigma(q,\rho)\geq 0}p^{(2)}(q,\rho) \ ,
\ee
while with $u\leq 0$  {\rm{(}}BCS repulsion{\rm{)}} it has the form
\be \label{lim-Var Princ-u negative} p:=\lim_\Lambda p_\sLambda
=\inf_{q\in\CC}\inf_{\rho\geq 0}p^{(2)}(q,\rho)=
\inf_{q\geq 0}\inf_{\rho\,:\,\sigma(q,\rho)\geq 0}p^{(2)}(q,\rho) \ .
\ee
\end{theorem}
Note that to obtain the approximating Hamiltonian
(\ref{Approx-2-eta=0}), the term $- {u}Q^*_\sLambda Q^\*_\sLambda
/{2V}$ in (\ref{hamiltonian}) is replaced by $- u(Q_\sLambda^* q+
Q^\*_\sLambda q^*)/2 + V u  |q|^2 / 2$ and ${v} N_\sLambda^2 /{2V}$
by $v\rho N_\sLambda - {V} v \rho^2 /2$.

We shall prove Theorem \ref{Main Theorem} in \textit{two} steps. Here we describe these steps for $u>0$
and before the end of the section we indicate the modifications necessary for the case $u\leq 0$.
\par
The first step which we call the \tit{first approximation} is to
\textit{linearize} the term $- {u}Q^*_\sLambda Q^\*_\sLambda /{2V}$
in $H_\sLambda$. For technical reasons we need to add to our
Hamiltonians some \textit{source terms}. Therefore, we define for
$\nu, \eta\in \CC$ \be \label{PBH-sources}
H_\sLambda(\nu,\eta):=H_\sLambda-(\nu Q^*_\sLambda +\nu^*
Q^\*_\sLambda)-{\sqrt V} \(\eta a^*_0 + \eta^* a^\*_0\) , \ee
and the \textit{first approximating} Hamiltonian
\begin{eqnarray}\label{Approx-1}
&&H^{(1)}_\sLambda(q,\nu,\eta):=T_\sLambda+\frac{v}{2V}N_\sLambda^2-\half
u(Q_\sLambda^* q+ Q^\*_\sLambda q^*) +\half V u  |q|^2 - \\
\nonumber && \hskip 2.7cm (\nu Q^*_\sLambda +\nu^* Q^\*_\sLambda)
-{\sqrt V} \(\eta a^*_0 + \eta^* a^\*_0\).
\end{eqnarray}
From (\ref{PBH-sources}) and (\ref{Approx-1}) we have
\begin{equation*}
H_\sLambda(\nu,\eta)=H^{(1)}_\sLambda(q,\nu,\eta)+H^r_\sLambda(q)
\end{equation*}
where
\be \label{Hrq}
H^r_\sLambda(q)= -
\frac{1}{2V}u(Q^*_\sLambda-V q^*)(Q^\*_\sLambda- V q) \leq 0.
\ee
First we show (see Section \ref{Section 1st Approx})  that with the
right choice of the parameter $q = \bar{q}$, the residual
perturbation $H^r_\sLambda(\bar{q})$ does not contribute to
$p_\sLambda(\nu,\eta)$, the pressure for the PBH (\ref{PBH-sources}) in the thermodynamic limit, i.e.,
the pressure corresponding to the Hamiltonian $H_\sLambda (\nu,\eta)$
coincides with the limit of $p^{(1)}_\sLambda(\bar{q},\nu,\eta)$,
the pressure for $H^{(1)}_\sLambda(\bar{q},\nu,\eta)$:
\begth
\label{Theor 1st Approx} For any $\nu$ and $\eta$ with
$|\nu|\leq 1$ and $|\eta|\leq 1$,
\be \label{lim 1st appr}
\lim_\Lambda  p_\sLambda(\nu,\eta)=\lim_\Lambda \sup_q p^{(1)}_\sLambda(q,\nu,\eta). \ee
In particular
\be
\lim_\Lambda  p_\sLambda(\eta)=\lim_\Lambda \sup_q p^{(1)}_\sLambda(q,\eta).
\ee
where
$p_\sLambda(\eta):=p_\sLambda(0,\eta)$ and $p^{(1)}_\sLambda(q,\eta):=p^{(1)}_\sLambda(q,0,\eta)$
are the pressures corresponding to the  Hamiltonians
$H_\sLambda(\eta):=H_\sLambda(0,\eta)$ and $H^{(1)}_\sLambda(q,\eta):=H^{(1)}_\sLambda(q,0,\eta)$
respectively.
\enth
Next, in Section \ref{Section 2nd Approx} we study a \textit{second
approximating} Hamiltonian obtained from (\ref{Approx-1}) by
replacing the term ${v}N_\sLambda^2/{2V}$ by a linear term $v \rho
N_\sLambda - {V}v\rho^2 /2$:
\be
\label{Approx-2}
H^{(2)}_\sLambda(q,\rho,\eta):=T_\sLambda+v\rho
N_\sLambda-\frac{1}{2}u(Q^*_\sLambda q + Q^\*_\sLambda q^*)
-\frac{V}{2}v\rho^2 +\frac{V}{2}u |q|^2 -{\sqrt V} \(\eta a^*_0
+ \eta^* a^\*_0\).
\ee
We denote the pressure corresponding to the
Hamiltonian (\ref{Approx-2}) by ${\tilde
p}^{(2)}_\sLambda(q,\rho,\eta)$. Note that by (\ref{Approx-2-eta=0})
and (\ref{Approx-2}) one has $H^{(2)}_\sLambda(q,\rho,0)=
H^{(2)}_\sLambda(q,\rho)$. We shall show in Lemma \ref{explicit p2}
that
\bes {\tilde
p}^{(2)}_\sLambda(q,\rho,\eta)=p^{(2)}_\sLambda(q,\rho) +|\eta|^2
\left\{\frac{f(0,\!\rho)-
|u||q|\cos(\theta-2\psi)}{f^2(0,\!\rho)-u^2|q|^2}\right\}
\ees
where
$\theta :=\arg q$ and $\psi :=\arg \eta$.
\par
Our next theorem establishes a similar
variational relation between the pressure $p_\sLambda(\eta)$ and
${\tilde p}^{(2)}_\sLambda(q, \rho,\eta)$:

\begin{theorem}
\label{Theor 2nd Approx}
\be
\label{lim 2nd appr} \lim_\Lambda p_\sLambda(\eta)
= \lim_\Lambda \sup_{q\in\CC}\inf_{\rho\geq 0}{\tilde p}^{(2)}_\sLambda(q, \rho,\eta)
= \lim_\Lambda \sup_{q\geq 0}\inf_{\rho\geq 0}p^{(2)}_\sLambda(q, \rho,\eta) \,,
\ee
where for $q\geq 0$ we put
\begin{equation}\label{press-real-q-eta}
p^{(2)}_\sLambda(q, \rho,\eta):={\tilde p}^{(2)}_\sLambda(qe^{{\rm i}(\pi +2\psi)},
\rho,\eta)
=p^{(2)}_\sLambda(q, \rho)+\frac{|\eta|^2}{f(0,\rho)-u q } \,.
\end{equation}
\end{theorem}
Note that the difference between the statement in Theorem \ref{Main
Theorem} and that in Theorem \ref{Theor 2nd Approx} (apart from the
$\eta$ dependence) is that the thermodynamic limit is taken
\textit{after} taking the \textit{infimum} over $\rho$ and
the \textit{supremum} over $q$. In the next theorem we show
that the order of the thermodynamic limit and taking the
\textit{infimum} and \textit{supremum} can be
reversed:
\begin{theorem}
\label{interchange Theor 2nd}
For $\eta\neq 0$,
\begin{equation}\label{lim-pres-eta}
p(\eta):=\lim_\Lambda p_\sLambda(\eta)=
\sup_{q\geq 0}\inf_{\rho\,:\,\sigma(q,\rho)\geq 0}p^{(2)}(q,\rho,\eta) \,,
\end{equation}
where we put
\begin{equation}\label{press-real-q-eta-lim}
p^{(2)}(q,\rho,\eta):=\lim_\Lambda p_\sLambda^{(2)}(q,\rho,\eta)=p^{(2)}(q,\rho)+
\frac{|\eta|^2}{f(0,\rho)-uq} \,,
\end{equation}
cf. expression (\ref{press-real-q-eta})\,.
\end{theorem}
In Lemma \ref{off source} we prove that $p=\lim_{\eta\to 0}p(\eta)$
so that Theorem \ref{interchange Theor 2nd} gives
\be
p=\lim_{\eta\to 0}\sup_{q\geq 0}\inf_{\rho\,:\,\sigma(q,\rho)\geq 0}p^{(2)}(q,\rho,\eta).
\ee
Finally in Lemma \ref{Var Princ} we
prove that the order of the limit $\eta\to 0$ and taking the
\textit{\textit{infimum}} and \textit{\textit{supremum}} can be
reversed to yield the main result Theorem \ref{Main Theorem} for the BCS attraction.
\par
The important difference for the \textit{repulsive} case, $u<0$, is that instead of (\ref{Hrq}) we now have
\be
\label{Hrq-w}
H^r_\sLambda(q)=
-\frac{1}{2V} u (Q^*_\sLambda-V q^*)(Q^\*_\sLambda- V q) \geq 0 \ .
\ee
Therefore the first approximation (Section \ref{Section 1st Approx}) should be constructed in the same
way as the second approximation (Section \ref{Section 2nd Approx}). The proof of the second part of
Theorem \ref{Main Theorem}, (\ref{lim-Var Princ-u negative}),
for $u \leq 0$ is given in Section \ref{Discussion} (f).

\par
It is important to note that the variational formula conjectured in
\cite{PZ-Pair} has the same Euler-Lagrange equations as those given
by Theorem \ref{Main Theorem}. Thus the detailed study of these
equations carried out in \cite{PZ-Pair} applies to our result.
In particular, this concerns the sequence of phase transitions in
the PBH model (\ref{hamiltonian}) and the conditions for the \textit{coexistence} of
the \textit{generalized} Bose condensation and the condensation of \textit{boson pairs},
see also Section \ref{Discussion}.
\par
The paper is organized as follows. We start by proving in Section
\ref{Superstab} that the PBH model (\ref{hamiltonian}) is
superstable. In Sections \ref{Section 1st Approx} and \ref{Section 2nd Approx} we shall assume that $u>0$.
Section \ref{Section 1st Approx} is devoted to establishing the
first approximation giving the proof of Theorem \ref{Theor 1st
Approx}. In Section \ref{Section 2nd Approx} we turn to the second
approximation giving the proof of Theorem \ref{Theor 2nd Approx} and
the other results needed to obtain Theorem \ref{Main Theorem} for $u > 0$.
Finally in Section \ref{Discussion} we discuss the variational
problem as well as related open questions for all values of $u$ and we finish the proof
of Theorem \ref{Main Theorem} for $u \leq 0$. Some commutator relations are given in
Appendix A and in Appendix B we give a bound needed in our proofs.
\section{Superstability}\label{Superstab}
In this section we establish the superstability of the PBH model
(\ref{hamiltonian}). When $u\leq 0$ superstability is obvious. To prove it for $u>0$ and $\alpha = v-u > 0$,
we shall need the following lemma
which is used in several other places in the paper.
\begin{lemma}
\label{Bounds on Q*Q}
The following inequality is satisfied
\be
Q^*_\sLambda Q^\*_\sLambda \leq N^2_\sLambda+MVN_\sLambda.
\label{B}
\ee
\end{lemma}
\begin{proof}
The inequalities
\begin{equation*}
\(\lambda^*(k)a_{k'}a^*_k \pm
\lambda^*(k')a^*_{-k'}a^\*_{-k}\)^*
\(\lambda^*(k)a_{k'}a^*_k \pm \lambda^*(k')a^*_{-k'}a^\*_{-k}\)\geq 0
\end{equation*}
and definition (\ref{A-oper}) imply that for $k\neq \{k',-k'\}$,
\bea
\label{estim-N-A-N}
&&-(N_k+|\lambda(k)|)N_{k'}-(N_{-k'}+|\lambda(k')|)N_{-k} \non\\
&&\hskip 1cm\leq -|\lambda(k)|^2(N_k+1)N_{k'}-|\lambda(k')|^2(N_{-k'}+1)N_{-k} \non\\
&& \hskip 2cm \leq \lambda^*(k)\lambda(k')A^*_k A_{k'}+\lambda^*(k')\lambda(k)A^*_{k'}A_k \\
&& \hskip 3cm \leq|\lambda(k)|^2(N_k+1)N_{k'}+|\lambda(k')|^2(N_{-k'}+1)N_{-k}\non\\
&& \hskip 4cm\leq (N_k+|\lambda(k)|)N_{k'}+(N_{-k'}+|\lambda(k')|)N_{-k} \ .\non
\eea
By (\ref{A-oper}) we also have
\begin{eqnarray}
\nonumber A^*_k A_k &=&N_kN_{-k}\ \ \ {\rm for}\ \ \ k\neq 0 \ , \\
A^*_0A_0&=&N_0(N_0-1)\leq N_0^2 \ .
\label{ineq-A-N}
\end{eqnarray}
Then by (\ref{estim-N-A-N}) and (\ref{ineq-A-N}) one
gets
\bea
Q^*_\sLambda Q^\*_\sLambda &= &{\hskip -0.5cm} \sum_{{k,
k'\in \Lambda^*,}\atop {k\neq k',\ k\neq -k'}}{\hskip -0.5cm}\lambda^*(k)\lambda(k')A^*_kA_{k'}
+2{\hskip -0.5cm}\sum_{k\in \Lambda^*,\ k\neq 0}
{\hskip -0.5cm}|\lambda(k)|^2A^*_k A_k +|\lambda(0)|^2A^*_0A_0\non\\
&=&
\half{\hskip -0.3cm}\sum_{{k, k'\in \Lambda^*,}\atop {k\neq k',\ k\neq -k'}}
{\hskip -0.5cm}\(\lambda^*(k)\lambda(k')A^*_k A_{k'}+\lambda^*(k')\lambda(k)A^*_{k'}A_k \)+
2{\hskip -0.5cm}\sum_{k\in \Lambda^*,\ k\neq 0}
{\hskip -0.5cm}|\lambda(k)|^2A^*_k A_k +|\lambda(0)|^2A^*_0A_0\non\\
&\leq&
\half{\hskip -0.3cm}\sum_{{k, k'\in \Lambda^*,}\atop {k\neq k',\ k\neq -k'}}{\hskip -0.5cm}
\((N_k+|\lambda(k)|)N_{k'}+(N_{-k'}+|\lambda(k')|)N_{-k}\)+
2{\hskip -0.5cm}\sum_{k\in \Lambda^*,\ k\neq 0}
{\hskip -0.5cm}N_kN_{-k}+N_0^2\non\\
&=& {\hskip -0.3cm}\sum_{{k, k'\in \Lambda^*,}\atop {k\neq k'}}
{\hskip -0.0cm} N_k N_{k'}+
{\hskip -0.5cm}+{\hskip -0.5cm}\sum_{k\in \Lambda^*,\ k\neq 0} {\hskip -0.0cm}N_kN_{-k}+N_0^2+
\sum_{{k, k'\in \Lambda^*,}\atop {k\neq k',\
k\neq -k'}}{\hskip -0.5cm}|\lambda(k)| N_{k'} \ .
\eea
Using the inequality
\be
N_k N_{-k}\leq \half\(N^2_k+N^2_{-k}\) \ ,
\ee
we get
\be
\sum_{k\in
\Lambda^*,\ k\neq 0} {\hskip -0.0cm}N_kN_{-k}\leq \sum_{k\in
\Lambda^*,\ k\neq 0} {\hskip -0.0cm}N_k^2.
\ee
Thus (\ref{B}) follows
by (\ref{N-Q-oper}) and (\ref{M}). \qquad \hfill $\square$

We now use the inequality (\ref{B}) in Lemma \ref{Bounds on Q*Q} to
prove superstability  of the model (\ref{hamiltonian}).
\begth\label{super} The Hamiltonian
(\ref{hamiltonian}) is superstable:
\be
H_\sLambda-\mu
N_\sLambda\geq T_\sLambda+\frac{1}{2V}\alpha N^2_\sLambda
-(\mu+R)N_\sLambda \label{D}
\ee
where $R:= Mu/2$ and $M$ is defined by (\ref{M}).
\enth
{\bf Proof:}\ \ From Lemma \ref{Bounds on Q*Q} \bea H_\sLambda-\mu
N_\sLambda &\geq & T_\sLambda+\frac{1}{2V}
\(v-u\)N^2_\sLambda-(\mu+R)N_\sLambda \non\\
& = & T_\sLambda+\frac{1}{2V}\alpha N^2_\sLambda -(\mu+R)N_\sLambda.
\label{D2}
\eea
Since we are assuming that $\alpha>0$, the estimate (\ref{D2}) implies superstability,
see \cite{Ru}.
\end{proof}
In the next two sections we develop the proofs for the variational formula for the pressure.
\section{The First Approximation}\label{Section 1st Approx}

Recall that the auxiliary Hamiltonians  $H_\sLambda(\nu,\eta)$ and
$H^{(1)}_\sLambda(q,\nu,\eta)$ are source dependent with $\nu, \eta\in \CC$, see (\ref{PBH-sources})
and (\ref{Approx-1}).
Since later we shall let $\nu$ and
$\eta$ tend to zero, we can assume that $|\nu|\leq 1$ and $|\eta|\leq 1$.
Because we are making the assumption on PBH (\ref{hamiltonian}) that $u>0$, it follows from (\ref{Hrq})
that $ H^r_\sLambda(q)\leq 0$.

Let $\nu \in \CC $ and $\phi:=\arg(\nu^*\lambda(k))$.
Then from
\begin{equation*}
(a^*_k \pm e^{-{\rm i}\phi}a^\*_{-k})(a^\*_k \pm e^{{\rm i}\phi}a^*_{-k})\geq 0
\end{equation*}
we get
\be
-|\nu|(N_k+N_{-k}+|\lambda(k)|)\leq \nu \lambda^*(k)A^*_k  +
\nu^* \lambda(k)A_k \leq |\nu|(N_k+N_{-k}+|\lambda(k)|)\ .
\label{K}
\ee
Also
\begin{equation*}
{\sqrt V} \(\eta a^*_0 + \eta^* a^\*_0\)= (a^*_0+{\sqrt V}\eta^*)(a^\*_0+{\sqrt V}\eta)-
a^*_0 a^\*_0-V|\eta|^2\geq-N_\sLambda-V|\eta|^2 \ .
\end{equation*}
Therefore, by Theorem \ref{super} one gets for $|\nu|\leq 1$ and $|\eta|\leq 1$,
the estimate:
\bea
H_\sLambda(\nu,\eta) -\mu N_\sLambda &\geq & H_\sLambda-\sum_{k\in \Lambda^*}(N_k+N_{-k}+
|\lambda(k)|)-N_\sLambda-V
-\mu N_\sLambda\non\\
&\geq & H_\sLambda-(\mu +3)N_\sLambda-\mathfrak{m}_\sLambda-V\non\\
&\geq & T_\sLambda+\frac{1}{4V}\alpha N^2_\sLambda -(\mu+3+R)N_\sLambda -(M+1)V \ .
\label{H}
\eea
Since $H^r_\sLambda(q)\leq 0$, we also have
\bea
H^{(1)}_\sLambda(q,\nu,\eta) -\mu N_\sLambda &\geq&
H_\sLambda(\nu,\eta) -\mu N_\sLambda \non\\
&\geq&  T_\sLambda+
\frac{1}{4V}\alpha N^2_\sLambda -(\mu+3+R)N_\sLambda - (M+1)V \ .
\label{I}
\eea
{\bf Proof of Theorem \ref{Theor 1st Approx}\,:}

For simplicity we shall prove this theorem for $\nu=0$. The proof
for a general $\nu$ follows through verbatim by translation for $\nu \neq 0$.
Clearly since $H_\sLambda^r\leq 0$, it follows from (\ref{I}) that
for any $q$ we have for the pressure  of the PBH (\ref{PBH-sources}) the estimate from below:
\begin{equation*}
p_\sLambda(\eta)\geq p^{(1)}_\sLambda(q,\nu=0,\eta)=p^{(1)}_\sLambda(q,\eta).\end{equation*}
Also for any $q$ one obviously has the estimate from above:
\begin{eqnarray*}
p_\sLambda (\eta) &=& p^{(1)}_\sLambda(q,\eta)+ \(p_\sLambda(\nu,\eta)-p^{(1)}_\sLambda(q,\nu,\eta))\)
\non\\
&&{\hskip 1cm}-\(p_\sLambda(\nu,\eta)-p_\sLambda(\eta)\)+
\(p^{(1)}_\sLambda(q,\nu,\eta))-p^{(1)}_\sLambda(q,\eta)\)\non\\
&\leq &\sup_{q'} p^{(1)}_\sLambda(q',\eta)+ \(p_\sLambda(\nu,\eta)-p^{(1)}_\sLambda(q,\nu,\eta)\)\non\\
&&{\hskip 1cm}-\(p_\sLambda(\nu,\eta)-p_\sLambda(\eta)\)+
\sup_{q'} \(p^{(1)}_\sLambda(\nu, q',\eta)-p^{(1)}_\sLambda(q',\eta)\),
\end{eqnarray*}
and, therefore, we get
\bea
\sup_q p^{(1)}_\sLambda(q,\eta)&\leq& p_\sLambda(\eta)\leq \sup_q p^{(1)}_\sLambda(q,\eta)
+ \inf_q \(p_\sLambda(\nu,\eta)-p^{(1)}_\sLambda(q,\nu,\eta)\)\non\\
&&{\hskip 1cm}-\(p_\sLambda(\nu,\eta)-p_\sLambda(\eta)\)+
\sup_q \(p^{(1)}_\sLambda(q,\nu,\eta)-p^{(1)}_\sLambda(q,\eta)\).
\eea
We shall prove in Lemma \ref{Easy} that, if $\nu_\sLambda \to 0 $ as $\Lambda \uparrow \mathbb{R}^\nu$, then
\be
\liminf_{\Lambda}(p_\sLambda(\nu_\sLambda,\eta )-p_\sLambda(\eta))=0 \ ,
\ee
and
\be
\limsup_{\Lambda}\{\sup_q(p^{(1)}_\sLambda(q,\nu_\sLambda,\eta)-p^{(1)}_\sLambda(q,\eta))\}=0 \ .
\ee
Next, with a \textit{particular choice} of $\nu_\sLambda$ that tends to zero as
$\Lambda \uparrow \mathbb{R}^\nu $,
we shall show also that
\be
\limsup_{\Lambda }\{\inf_q (p_\sLambda(\nu_\sLambda,\eta)-
p^{(1)}_\sLambda(q,\nu_\sLambda,\eta))\}=0 \ .
\ee
This last result (which is proved in Lemma \ref{hard}) is much harder and requires the
arguments developed in \cite{BBZKT}.
Putting these together we get
\be
\lim_\Lambda  p_\sLambda(\eta)=\lim_\Lambda \sup_q p^{(1)}_\sLambda(q,\eta)\,,
\ee
that proves Theorem \ref{Theor 1st Approx}\,. \qquad \hfill $\square$

We now prove the two lemmas quoted earlier.
\blem \label{Easy}
\be
\label{Easy-1}
\liminf_{\Lambda }(p_\sLambda(\nu_\sLambda,\eta )-p_\sLambda(\eta))= 0
\ee
and
\be
\label{Easy-2}
\limsup_{\Lambda }(p^{(1)}_\sLambda(q,\nu_\sLambda ,\eta)-p^{(1)}_\sLambda(q,\eta))=0
\ee
\elem
\begin{proof} Writing $\nu = x+iy$, using the convexity of the pressure and (\ref{K}) we get
\bea
p_\sLambda(\nu,\eta)-p_\sLambda(\eta)
&\geq & x \(\frac{\partial}{\partial x}p_\sLambda(\nu,\eta)\)\Bigg|_{\nu=0}
+y\(\frac{\partial}{\partial y}p_\sLambda(\nu,\eta)\)\Bigg|_{\nu=0}\non \\
&=&
\frac{1}{V} \<\nu Q^*_\sLambda +\nu^* Q^\*_\sLambda\>_{H_\sLambda(\eta) }\non\\
&\geq&
-\frac{1}{V} |\nu|\sum_{k \in \Lambda^*} \<N_k+N_{-k}+|\lambda(k)|\>_{H_\sLambda(\eta) }\non\\
&\geq& -\frac{1}{V} |\nu|\(2\<N_\sLambda\>_{H_\sLambda(\eta)}+\mathfrak{m}_\sLambda\)\geq -K |\nu| \ ,
\eea
by (\ref{M}) and Lemma \ref{App1}.
Therefore if $\nu_\sLambda \to 0$ as $\Lambda \uparrow  \mathbb{R}^\nu $, we get (\ref{Easy-1}):
\be
\liminf_{\Lambda }(p_\sLambda(\nu_\sLambda,\eta )-p_\sLambda(\eta))=0 \ .
\ee
Similarly one gets
\be
\sup_q\(p^{(1)}_\sLambda(q,\nu,\eta)-p^{(1)}_\sLambda(q,\eta)\)
\leq
\frac{1}{V} |\nu|\sup_q\(2\<N_\sLambda\>_{H^0_\sLambda (q,\nu,\eta)}+\mathfrak{m}_\sLambda\)
\leq K |\nu| \ ,
\ee
by (\ref{M}), (\ref{I}) and Lemma \ref{App1}.
Thus
\be
\limsup_{\Lambda}\{\sup_q(p^{(1)}_\sLambda(q,\nu_\sLambda ,\eta)-p^{(1)}_\sLambda(q,\eta))\}=0 \ ,
\ee
that implies (\ref{Easy-2}).
\end{proof}
\blem \label{hard} There exists a sequence
$\left\{\nu_\sLambda\right\}_\sLambda$  that tends to $0$ as
$\Lambda \uparrow \mathbb{R}^\nu $, such that
\be
\label{1st-approx}
\limsup_{\Lambda }\inf_q \(p_\sLambda(\nu_\sLambda,\eta)-
p^{(1)}_\sLambda(q,\nu_\sLambda, \eta)\)=0 \,.
\ee
\elem
\begin{proof}
Using the Bogoliubov convexity inequality \cite{BBZKT}:
\be
\frac{\mbox{Tr} (A-B)e^B}{\mbox{Tr} e^B}\leq \ln \mbox{Tr} e^A-
\ln \mbox{Tr} e^B \leq \frac{\mbox{Tr} (A-B)e^A}{\mbox{Tr} e^A}
\label{J}
\ee
and (\ref{Hrq}) we get the estimate
\begin{equation*}
0 \leq p_\sLambda(\nu,\eta)-p^{(1)}_\sLambda(q,\nu,\eta) \leq
\frac{1}{2V^2} u
\<(Q^*_\sLambda-Vq^*)(Q^\*_\sLambda-Vq)\>_{H_\sLambda (\nu,\eta)} .
\end{equation*}
Let
$\delta Q^\*_\sLambda (\nu,\eta):=Q^\*_\sLambda -\<Q^\*_\sLambda\>_{H_\sLambda(\nu,\eta)}$ and
let
\[
\Delta_\sLambda(\nu,\eta):=
 \<\delta Q^*_\sLambda (\nu,\eta)\ \delta Q^\*_\sLambda (\nu,\eta)\>_{H_\sLambda(\nu,\eta)} \geq 0.
\]
Then
\be
\label{est-inf-1st-approx}
\inf_q \(p_\sLambda(\nu,\eta)-p^{(1)}_\sLambda(q,\nu,\eta)\) \leq
\frac{u}{2V^2}\Delta_\sLambda(\nu,\eta) \ .
\ee
We want to obtain an estimate for $ \Delta_\sLambda(\nu,\eta)$ in terms of
$\nu$ and $V$.
\par
Let
\be
D_\sLambda(\nu,\eta):=  \(\delta Q^*_\sLambda (\nu,\eta),\
\delta Q^\*_\sLambda (\nu,\eta)\)_{H_\sLambda(\nu,\eta)} ,
\ee
where
$(\cdot\ ,\ \cdot)_H$ denotes the Bogoliubov-Duhamel \textit{inner
product} with respect to the Hamiltonian $H$, see for example
\cite{BBZKT} or \cite{Gi}. Using the Ginibre
inequality  (e.g. (2.10) in \cite{Gi}) we get
\bea
\label{Gin-ineq}
\non &&\Delta_\sLambda(\nu,\eta)\leq\frac{1}{2} \<\delta Q^*_\sLambda
(\nu,\eta)\ \delta Q^\*_\sLambda (\nu,\eta)+\delta Q^*_\sLambda
(\nu,\eta)\
\delta Q^\*_\sLambda (\nu,\eta)\>_{H_\sLambda(\nu,\eta)}\\
\non &&\leq D_\sLambda(\nu,\eta)+\half
\left\{\beta D_\sLambda(\nu,\eta) \right\}^{1/2}\left\{\< [Q^*_\sLambda,\
[H_\sLambda(\nu,\eta)-\mu N_\sLambda , Q^\*_\sLambda]]\>_{H_\sLambda(\nu,\eta)}
\right\}^{1/2}.
\eea
We shall show in Appendix A that there is a real number $C$ such that
\begin{equation*}
\< [Q^*_\sLambda,\
[H_\sLambda(\nu,\eta)-\mu N_\sLambda,Q^\*_\sLambda]]\>_{H_\sLambda(\nu,\eta)} \leq
C\,V^{3/2} \ .
\end{equation*}
Thus \be \label{C} \Delta_\sLambda(\nu,\eta)\leq
D_\sLambda(\nu,\eta)+(C\beta)^{1/2}
\left\{V^{3/2}D_\sLambda(\nu,\eta) \right\}^{1/2}. \ee From the
definition of the Bogoliubov-Duhamel inner product we have
\begin{equation*}
D_\sLambda(\nu,\eta)= V  \frac{\partial^2}{\partial \nu \partial
\nu^*}p_\sLambda(\nu,\eta) = \frac{V}{4}
\left\{\frac{\partial^2}{\partial x^2 }+ \frac{\partial^2}{\partial
y^2}\right\}p_\sLambda(\nu,\eta) \ .
\end{equation*}
Here we consider the pressure
$p_\sLambda(\nu,\eta)$ as a
function of two real variables, $x= \mathfrak{Re}\,\nu$ and $y= \mathfrak{Im}\,\nu$.
Since $u>0$, then following the \textit{Approximating Hamiltonian Method}
for \textit{attractive} interactions  \cite{BBZKT} we consider the
integral
\begin{equation*}
I_\Lambda (\delta):=
\int_{[-\delta,\delta]^2}dx\,dy\frac{\partial^2}{\partial
x^2}p_\sLambda(\nu,\eta) \,.
\end{equation*}
With $\nu_+ := \delta + i y$ and $\nu_- := -\delta + i y$, this
integral is equal to
\begin{eqnarray*}
I_\Lambda (\delta) &=& \int_{[-\delta,\delta]} dy\left \{\frac{\partial}{\partial
x}p_\sLambda(\nu_+,\eta)
-\frac{\partial}{\partial x}p_\sLambda(\nu_-,\eta)\right \}\non\\
&=& \frac{1}{V}
\int_{[-\delta,\delta]}dy\left \{\<Q^\*_\sLambda + Q^*_\sLambda\>_{H_\sLambda(\nu_+,\eta)}
-\<Q^\*_\sLambda + Q^*_\sLambda\>_{H_\sLambda(\nu_-,\eta)}\right \}.
\end{eqnarray*}
Then by (\ref{K}) one gets
\begin{equation*}
|I_\Lambda
(\delta)|\leq \frac{2}{V} \int_{[-\delta,\delta]}dy\left \{\<{\tilde
N}_\sLambda \>_{H_\sLambda(\nu_+,\eta)} +\<{\tilde N}_\sLambda
\>_{H_\sLambda(\nu_-,\eta)}\right \} \,,
\end{equation*}
where ${\tilde N}_\sLambda
:=\sum_{k \in \Lambda^*}(N_k+N_{-k}+|\lambda(k)|)/2$. Since by (\ref{H})
and Lemma \ref{App1}, the expectation $\<N_\sLambda/V \>_{H_\sLambda(\nu,\eta)}$ is
bounded uniformly in $\nu$ and in $V$, we obtain the estimate
\begin{eqnarray*}
\nonumber && \left |\int_{[-\delta,\delta]^2}dx\,dy
\frac{\partial^2}{\partial x^2}p_\sLambda(\nu,\eta)\right| \leq \\
&& \frac{2}{V}\int_{[-\delta,\delta]}dy \left \{\<N_\sLambda
\>_{H_\sLambda(\nu_+,\eta)}
+\<N_\sLambda\>_{H_\sLambda(\nu_-,\eta)}+ \mathfrak{m}_\sLambda \right
\}
\leq 2 \tilde{C} \delta \,.
\end{eqnarray*}
Similarly one gets the estimate
\begin{equation*}
\left |\int_{[-\delta,\delta]^2}dx\,dy \frac{\partial^2}{\partial
y^2}p_\sLambda(\nu,\eta)\right|\leq 2 \tilde{C} \delta \,.
\end{equation*}
These give
\be
\label{ineq-D-1}
\int_{[-\delta,\delta]^2}dx\,dy D_\sLambda(\nu,\eta)\leq \tilde{C} V \delta \,.
\ee
Since the integrand is continuous, by the \textit{integral mean-value}
theorem there exists a sequence
$\left\{\nu_\sLambda\right\}_\sLambda$ with $|\nu_\sLambda | \leq
\delta$ such that
\begin{equation*}
\int_{[-\delta,\delta]^2}dx\,dy
D_\sLambda(\nu,\eta)=(2\delta)^2D_\sLambda(\nu_\sLambda,\eta) \,.
\end{equation*}
The last equation and inequality (\ref{ineq-D-1}) imply that
\begin{equation*}
D_\sLambda(\nu_\sLambda,\eta)\leq \frac{\tilde{C} V }{4\delta} \,,
\end{equation*}
which together with (\ref{C}) give the estimate
\begin{equation*}
\frac{1}{V^2}\,\,\Delta_\sLambda(\nu_\sLambda,\eta)\leq
\frac{\tilde{C}}{4 V\delta}+\frac{(\tilde{C} C \beta)^{1/2}}{2V^{3/4}\delta^{1/2}} \, .
\end{equation*}
Choosing $\delta=\delta_\sLambda$ such that $\delta_\sLambda
\rightarrow 0$, but $\ V \, \delta_\sLambda \to\infty$, we get
\begin{equation*}
\lim_{\Lambda}\frac{1}{V^2}\,\,\Delta_\sLambda(\nu_\sLambda,\eta)=0 \, .
\end{equation*}
By (\ref{est-inf-1st-approx}) this completes the proof
of the lemma.
\end{proof}
This proves the first
approximation. In the next section we deal with the second one.
\section{The Second Approximation}\label{Section 2nd Approx}

Note that from definitions (\ref{Approx-1}) and (\ref{Approx-2}) of
the \textit{first} and the \textit{second} approximating
Hamiltonians, $H^{(1)}_\sLambda(q,\nu,\eta)$ and
$H^{(2)}_\sLambda(q,\rho,\eta)$, respectively, it follows that
\be \label{1st-2d Ham}
H^{(1)}_\sLambda(q,\nu=0,\eta) -
H^{(2)}_\sLambda(q,\rho,\eta)=\frac{1}{2V}\, v
(N_\sLambda-\rho)^2\geq 0 \, .
\ee
Later in this section we shall
show (see Lemma \ref{explicit p2} and Remark \ref{arg of q}) that
\begin{equation}\label{estim-Rem4.1}
{\tilde p}^{(2)}_\sLambda(q,\rho,\eta)\leq {\tilde p}^{(2)}_\sLambda(|q|e^{{\rm i}(\pi+2\psi)},\rho,\eta)
=p^{(2)}_\sLambda(|q|,\rho,\eta) \, .
\end{equation}
In Lemma \ref{inf p2 rho} we prove
that for each $q\geq 0$ there is a unique density $\rho={\bar \rho}_\sLambda(q,\eta)>0$,
such that
\be \label{pressure2 inf}
p^{(2)}_\sLambda(q, {\bar \rho}_\sLambda(q,\eta),\eta)=\inf_\rho
p^{(2)}_\sLambda(q,\rho,\eta) \,.
\ee
We can also show (Lemma \ref{sup p2 q}) that there is at
least one $q = {\bar q}_\sLambda (\eta)>0$,  such that
\be \label{pressure2 sup-inf}
p^{(2)}_\sLambda({\bar q}_\sLambda, {\bar \rho}_\sLambda({\bar
q}_\sLambda),\eta) =\sup_q p^{(2)}_\sLambda(q, {\bar
\rho}_\sLambda(q),\eta)=\sup_q\inf_\rho p^{(2)}_\sLambda(q,
\rho,\eta) \,.
\ee
For the sake of simplicity below we shall omit the variable $\eta $, and we
put
\begin{equation*}
{\bar \rho}_\sLambda(q,\eta):= {\bar \rho}_\sLambda(q)\,\,\,\,\,\,\, \mbox{and} \,\,\,\,\,\,\,
{\bar q}_\sLambda(\eta):= {\bar q}_\sLambda \,.
\end{equation*}
Finally, we shall show in Lemma \ref{L} that if $\eta\neq 0$, then
\be \label{diff}
\lim_\Lambda \{p^{(2)}_\sLambda({\bar q}_\sLambda,
{\bar \rho}_\sLambda({\bar q}_\sLambda),\eta)
-p^{(1)}_\sLambda({\bar q}_\sLambda e^{{\rm i}(\pi+2\psi)},\eta)\}=0 \,.
\ee
We start by proving Theorem \ref{Theor 2nd Approx}, assuming the
results of Lemmas \ref{explicit p2} - \ref{L}, which
we prove later.

{\bf Proof of Theorem \ref{Theor 2nd Approx}\,:}

We have to prove the limit (\ref{lim 2nd appr}) i.e. that
\be \label{Theor 2nd Approx 2}
p(\eta):=\lim_\Lambda p_\sLambda(\eta)
= \lim_\Lambda p^{(2)}_\sLambda({\bar q}_\sLambda, {\bar \rho}_\sLambda({\bar q}_\sLambda),\eta)\,.
\ee
First, by (\ref{1st-2d Ham}) and  (\ref{estim-Rem4.1}) we have for all values of the variational
parameters $q$, $\rho$ and the source parameter $\eta$ that
\begin{equation*}
p^{(1)}_\sLambda(q,\eta):=p^{(1)}_\sLambda(q,\nu=0,\eta)\leq {\tilde p}^{(2)}_\sLambda(q,\rho,\eta)
\leq p^{(2)}_\sLambda(|q|,\rho,\eta) \, .
\end{equation*}
Therefore,
\begin{equation*}
p^{(1)}_\sLambda(q,\eta)\leq \inf_\rho p^{(2)}_\sLambda(|q|,\rho,\eta)
=p^{(2)}_\sLambda(|q|, {\bar \rho}_\sLambda(|q|),\eta)\,\end{equation*}
and thus by definition  (\ref{press-real-q-eta}) we obtain
\begin{equation*}
\sup_q p^{(1)}_\sLambda(q,\eta)\leq \sup_q p^{(2)}_\sLambda(|q|, {\bar \rho}_\sLambda(|q|),\eta)
=\sup_{q\geq 0}p^{(2)}_\sLambda(q, {\bar \rho}_\sLambda(q),\eta)
=p^{(2)}_\sLambda({\bar q}_\sLambda, {\bar \rho}_\sLambda({\bar q}_\sLambda),\eta) \, .
\end{equation*}
This estimate implies that
\be
\lim_\Lambda \sup_q p^{(1)}_\sLambda(q,\eta) \leq
\lim_\Lambda p^{(2)}_\sLambda({\bar q}_\sLambda, {\bar \rho}_\sLambda({\bar q}_\sLambda),\eta) \,.
\label{upper}
\ee
On the other hand for all $\eta$ we obviously have
\begin{eqnarray}\label{lower1}
\sup_q p^{(1)}_\sLambda(q,\eta)\geq p^{(1)}_\sLambda({\bar q}_\sLambda e^{{\rm i}(\pi+2\psi)},\eta)
&=& p^{(2)}_\sLambda({\bar q}_\sLambda, {\bar \rho}_\sLambda({\bar q}_\sLambda),\eta) \\
\nonumber &-&\(p^{(2)}_\sLambda({\bar q}_\sLambda, {\bar \rho}_\sLambda({\bar q}_\sLambda), \eta)
-p^{(1)}_\sLambda({\bar q}_\sLambda e^{{\rm i}(\pi+2\psi)},\eta)\).
\end{eqnarray}
Now the limit (\ref{diff}) and the estimate (\ref{lower1}) imply that
\begin{equation} \label{lower}
\lim_\Lambda
\sup_q p^{(1)}_\sLambda(q,\eta)\geq \lim_\Lambda
p^{(2)}_\sLambda({\bar q}_\sLambda, {\bar \rho}_\sLambda({\bar
q}_\sLambda),\eta).
\end{equation}
Taking into account (\ref{upper}) and (\ref{lower}) we get
\begin{equation*}
\lim_\Lambda \sup_q
p^{(1)}_\sLambda(q,\eta)
= \lim_\Lambda p^{(2)}_\sLambda({\bar q}_\sLambda, {\bar \rho}_\sLambda({\bar q}_\sLambda),\eta)\,.
\end{equation*}
Combining this result with Theorem \ref{Theor 1st Approx} we get
(\ref{Theor 2nd Approx 2}), i.e. the proof of  Theorem \ref{Theor 2nd Approx}.
\qquad \hfill $\square$

Now we return to proof of the lemmas quoted earlier.
\blem \label{explicit p2}
Let the functions $f$ and $h$  and the
spectral function
$E(k,\!q,\!\rho)$ be as defined in (\ref{fh}) and (\ref{E}) respectively.\\
(i) If $f(0,\rho)>u|q|\geq 0$, the pressure
${\tilde p}^{(2)}_\sLambda(q,\rho,\eta)$ corresponding to
$H^{(2)}_\sLambda(q,\rho,\eta)$ is given by
\bea
\label{diag p}
{\tilde p}^{(2)}_\sLambda(q,\rho,\eta)&=&-\frac{1}{\beta V}\sum_{k\in
\Lambda^*}\ln \{1-\exp(-\beta E(k,\!q,\!\rho))\}-\frac{1}{2V}
\sum_{k\in \Lambda^*}\(E(k,\!q,\!\rho) -f(k,\!\rho)\)\non \\
&+& |\eta|^2 \left\{\frac{f(0,\!\rho)- |u q|\cos(\theta-2\psi)}{f^2(0,\!\rho)-u^2|q|^2}\right\}
-\frac{1}{2}u|q|^2 +\frac{1}{2}v\rho^2 \, ,
\eea
where $\theta =\arg q$ and $\psi =\arg \eta $ .

(ii) If $f(0,\rho)\leq u|q|$, then ${\tilde p}^{(2)}_\sLambda(q,\rho,\eta)$ is infinite.
\elem
\bpr
(i) By  (\ref{fh}) and (\ref{Approx-2}) we can write
$H^{(2)}_\sLambda(q,\rho,\eta)-\mu N_\sLambda$ in the form
\begin{eqnarray*}
H^{(2)}_\sLambda(q,\rho,\eta)-\mu N_\sLambda &=& \sum_{k\in \Lambda^*}
\{f(k,\!\rho) a^*_ka^\*_k -\half \(h(k,\!q) a^*_ka^*_{-k}+ h^*(k,\!q) a^\*_{-k}a^\*_k \)\}\non \\
&-& {\sqrt V} \(\eta a^*_0 + \eta^* a^\*_0\)+V W(q,\!\rho)\,,
\end{eqnarray*}
where
\begin{equation*}
W(q,\!\rho)=\frac{1}{2}u |q|^2 -\frac{1}{2}v\rho^2 \, .
\end{equation*}
Let $q \lambda^{*}(k)=|q \lambda^{*}(k)|e^{{\rm i}\theta(k)}$.
Then with $a^\*_k={\tilde a}_k e^{{\rm i}
{\theta(k)}/{2}}$ ,
for $k\in \Lambda^*$, one gets
\begin{eqnarray}
H^{(2)}_\sLambda(q,\rho,\eta)-\mu N_\sLambda
&=& \sum_{k\in \Lambda^*}\{f(k,\!\rho) {\tilde a}_k^*{\tilde a}_k -
\half |h(k,\!q)|\( {\tilde a}_k^*{\tilde a}^*_{-k}+{\tilde
a}_{-k}{\tilde a}_k \)\}
\non\\
&-& {\sqrt V} \(\eta e^{-{\rm i} {\theta}/{2}} {\tilde a}^*_0 + \eta^*
e^{{\rm i} {\theta}/{2}}{\tilde a}_0\)+ V W(q,\!\rho)\,,\label{bilinear}
\end{eqnarray}
where $\theta = \arg q=\theta(0)$.\\
Note that if $f(0,\rho)>u|q|\geq 0$, then $f(k,\rho)>|h(k,q)|\geq 0$ for all $k\in \Lambda^*$,
so that $E(k,\!q,\!\rho)$ is well-defined and \textit{positive}, see (\ref{E}). Let
\be
x^2_k=\half \left\{\frac{f(k,\!\rho)}{E(k,\!q,\!\rho)}+1\right\} \ \ \ \
{\rm and}\ \ \ \ y^2_k=
\half \left\{\frac{f(k,\!\rho)}{E(k,\!q,\!\rho)}-1\right\}.
\ee
Then the \textit{canonical Bogoliubov} transformation:
${\tilde a}^\*_k=x_k\alpha^\*_k -y_k\alpha^*_{-k}$, gives
\bea
\label{diag}
&& H^{(2)}_\sLambda(q,\rho,\eta)-\mu N_\sLambda =
\sum_{k\in\Lambda^*} E(k,\!q,\!\rho)\alpha^*_k \alpha^\*_k
-{\sqrt V} \(\xi \alpha^*_0 + \xi^* \alpha_0\)\non\\
&&\, + \, \half \sum_{k\in \Lambda^*}\(E(k,\!q,\!\rho) -f(k,\!\rho)\)
+VW(q,\!\rho)\,,
\eea
where $\alpha^*_k$ and $\alpha^\*_k$ , $k\in \Lambda^*$, are boson creation and annihilation
operators and
\begin{equation*}
\xi=\eta \ x_0 e^{-{\rm i} {\theta}/{2}}  - \eta^* y_0 e^{{\rm i} {\theta}/{2}} \ .
\end{equation*}
We note that
\begin{equation*}
|\xi|^2={|\eta|^2} \, \frac{f(0,\!\rho)- |u q|\cos(
\theta - 2\psi)}{E(0,\!q,\!\rho) } \ \ .
\end{equation*}
From the diagonal form of $H^{(2)}_\sLambda(q,\rho,\eta)-\mu
N_\sLambda$ in (\ref{diag}) we get the pressure (\ref{diag p}).\\
(ii) Now let $f(0,\rho) < u|q|$. Then the quadratic Hamiltonian (\ref{bilinear})
is not bounded from below.
This means that the trace in (\ref{press-main}) is divergent and therefore the pressure
${\tilde p}^{(2)}_\sLambda(q,\rho,\eta)$ is infinite. If $f(0,\rho) = u|q|$, then by definitions (\ref{fh})
and the conditions on ${\tilde \lambda}(k)$ at least the zero-mode term of the
Hamiltonian (\ref{bilinear}) is \textit{not} positive. This again implies that the
trace in expression (\ref{press-main}) diverges.
\epr
\begin{remark} \label{arg of q}
From the explicit formula (\ref{diag p}) it follows that
\begin{equation*}
{\tilde p}^{(2)}_\sLambda(q,\rho,\eta)
\leq  {\tilde p}^{(2)}_\sLambda(|q|e^{{\rm i}(\pi+2\psi)},\rho,\eta)
=p^{(2)}_\sLambda(|q|,\rho,\eta)\,.
\end{equation*}
Recall that by (\ref{press-real-q-eta}) and  (\ref{diag p}) one gets for $q\geq 0$
\begin{eqnarray}
\nonumber p^{(2)}_\sLambda(q,\rho,\eta)&=&-\frac{1}{\beta V}\sum_{k\in
\Lambda^*}\ln \{1-\exp(-\beta E(k,\!q,\!\rho))\}-\frac{1}{2V}
\sum_{k\in \Lambda^*}\(E(k,\!q,\!\rho) -f(k,\!\rho)\) \\
&+& \frac{|\eta|^2}{f(0,\!\rho)-u q}
-\half uq^2 +\half v\rho^2 \,.\label{diag p-real}
\end{eqnarray}
\end{remark}
\blem \label{inf p2 rho}
Let $\eta \neq 0$. Then there are numbers $0<{\tilde
\rho}_1(q,\!\eta)<{\tilde \rho}_2(q,\!\eta)<\infty$, such that the
\textit{\textit{infimum}} of $p^{(2)}_\sLambda(q,\rho,\eta)$ over
$\rho$ is attained in the interval  $({\tilde \rho}_1(q,\!\eta),{\tilde
\rho}_2(q,\!\eta))$ and if ${\bar\rho}_\sLambda(q)$ is a value of
$\rho$ at which the \textit{\textit{infimum}} is attained, then
${\ds {\partial p^{(2)}_\sLambda}(q,{\bar\rho}_\sLambda(q),\eta)/{\partial  \rho
}=0}$. Moreover, if $0<q_0<\infty$, then
\begin{equation*}
\inf_{q\leq q_0}(v{\tilde \rho}_1(q,\!\eta)-(\mu+uq)_+)>0 \ \ \,\,\,\,\,\,\,\,\,\,
{\rm{and}}  \ \ \,\,\,\,\,\,\,\,\,\, \sup_{q\leq q_0}{\tilde \rho}_2(q,\!\eta)<\infty \ ,
\end{equation*}
where $s_\pm:=\max(0,\pm s)$ for $s\in\RR$.
\elem
\bpr
By (\ref{diag p-real}) we have
\bea
&&\frac{\partial p^{(2)}_\sLambda}{\partial  \rho }(q,\rho,\eta)
=-\frac{v}{V}\sum_{k\in \Lambda^*}\left \{\ \frac{1}{\exp(\beta E(k,\!q,\!\rho))-1}\
\frac{f(k,\!\rho)}{E(k,\!q,\!\rho)}
+\frac{1}{2}\(\frac{f(k,\!\rho)}{E(k,\!q,\!\rho)}-1\)\right\}\non \\
&&\hskip 6cm-\frac{v|\eta|^2}{(f(0,\!\rho)-uq)^2} + v\rho \ .
\label{partial rho} \eea
From (\ref{partial rho}) we get
\begin{equation*}
\frac{\partial
p^{(2)}_\sLambda}{\partial  \rho }(q,\rho,\eta) \leq
-\frac{v|\eta|^2}{(f(0,\!\rho)-uq)^2}+v\rho \ .
\end{equation*}
Let
$x:=v\rho-(\mu+uq)_+$.
Using the identity $\mu+u q =(\mu+uq)_+ -(\mu+uq)_-  $ we obtain
\begin{equation*}
\frac{\partial
p^{(2)}_\sLambda}{\partial  \rho }(q,\rho,\eta) \leq
-\frac{v|\eta|^2}{((\mu+uq)_- +x)^2}+(\mu+uq)_+ +x \ .
\end{equation*}
As $x \to 0$, the right-hand side of the last inequality becomes negative.
Therefore, there exists $\delta(q,\!\eta)>0$ such that the
\textit{\textit{infimum}} of $p^{(2)}_\sLambda(q,\rho,\eta)$ over
$\rho$ cannot be achieved if $v\rho-(\mu+uq)_+<\delta(q,\!\eta)$, i.e.
$\rho<{\tilde \rho}_1(q,\!\eta):=((\mu+uq)_+ +\delta(q,\!\eta))/v$.
\par
It is clear that if $0<q_0<\infty$, then $\inf_{q\leq q_0}\delta(q,\!\eta)>0$.
\par
Suppose now that $\rho>{\tilde \rho}_1(q,\!\eta)$ and take $v\rho>\max(2\mu,2q+2)$.
Then for $k\in \Lambda^*$ one has
$E(k,\!q,\!\rho)>\max(\eps(k),1)$.
Therefore, using
\bes
0\leq \frac{f(k,\!\rho)}{E(k,\!q,\!\rho)}-1\leq \frac{|h(k,q)|}{E(k,\!q,\!\rho)}\leq uq|\lambda(k)|,
\ees
we obtain the estimate
\bea
\frac{\partial p^{(2)}_\sLambda}{\partial  \rho }(q,\rho,\eta)
&=&-\frac{v}{V}\sum_{k\in \Lambda^*}\left \{\ \frac{1}{\exp(\beta E(k,\!q,\!\rho))-1}\
+\half\coth\half\beta E(k,\!q,\!\rho)\(\frac{f(k,\!\rho)}{E(k,\!q,\!\rho)}-1\)\right\}\non \\
&-&\frac{v|\eta|^2}{(f(0,\!\rho)-uq)^2} + v\rho  \label{partial rho2} \\
&\geq &-\frac{v}{V}\sum_{k\in \Lambda^*}\ \frac{1}{\exp[\beta \max(\eps(k),1)]-1} \
-\frac{v}{2V}\, uq \ \sum_{k\in \Lambda^*}|\lambda(k)|-\frac{v|\eta|^2}{\delta(q,\!\eta)^2} + v\rho \non \ .
\eea
Making use of (\ref{M}), this implies that there exists a volume $V_0$ independent of $q$ and
$\rho$,  and $K(q,\!\eta)>0$ such that if $V>V_0$, then
\begin{equation*}
\frac{\partial
p^{(2)}_\sLambda}{\partial  \rho }(q,\rho,\eta)
\geq-K(q,\!\eta)+v\rho \,,
\end{equation*}
and therefore, if $\rho$ is large enough, then
${\ds \frac{\partial p^{(2)}_\sLambda}{\partial  \rho
}(q,\rho,\eta)>0}$. As a consequence, there is ${\tilde
\rho}_2(q,\!\eta)$ such that the \textit{\textit{infimum}} of
$p^{(2)}_\sLambda(q,\rho,\eta)$ is attained in the interval $({\tilde
\rho}_1(q,\!\eta),{\tilde \rho}_2(q,\!\eta))$. If
${\bar\rho}_\sLambda(q)$ is a value of $\rho$ at which the
\textit{\textit{infimum}} is attained, then ${\ds\frac{\partial
p^{(2)}_\sLambda}{\partial \rho
}(q,{\bar\rho}_\sLambda(q),\eta)=0}$.\\
Let $0<q_0<\infty$. Then one can see that $\sup_{q\leq q_0}K(q,\!\eta)<\infty$,
and therefore we get $\sup_{q\leq q_0}{\tilde \rho}_2(q,\!\eta)<\infty$.
\epr
\blem \label{sup p2 q}  Let $\eta\neq0$. Then there is $q_0(\eta)<\infty$ such that the
\textit{\textit{supremum}} of
$p^{(2)}_\sLambda(q,{\bar\rho}_\sLambda(q),\eta)$ with respect to
$q$  is attained in the interval $(0, q_0(\eta))$ for all $\Lambda$ and if ${\bar
q}_\sLambda$ is a maximizer of
$p^{(2)}_\sLambda(q,{\bar\rho}_\sLambda(q),\eta)$, then
\begin{equation*}
\ds \frac{dp^{(2)}_\sLambda}{d q}({\bar q}_\sLambda,{\bar\rho}_\sLambda({\bar q}_\sLambda),\eta)=0 \, .
\end{equation*}
There exists ${\bar c}_0(\eta)$ such that
for all $\Lambda$
\begin{equation*}
f(0,\!{\bar\rho}_\sLambda({\bar q}_\sLambda))-u{\bar q}_\sLambda>{\bar c}_0(\eta)\, ,
\end{equation*}
if ${\bar q}_\sLambda$ is a maximizer of
$p^{(2)}_\sLambda(q,{\bar\rho}_\sLambda(q),\eta)$.
\elem
\bpr
Recall that $v-u:=\alpha>0 $.
Differentiating $p^{(2)}_\sLambda(q,\rho,\eta)$ we get
\bea
&& \frac{\partial p^{(2)}_\sLambda}{\partial q}(q,\rho,\eta)
=\frac{u^2q}{V}\sum_{k\in \Lambda^*}|\lambda(k)|^2\left \{\ \frac{1}{\exp(\beta E(k,\!q,\!\rho))-1}\
\frac{1}{E(k,\!q,\!\rho)}+  \frac{1}{2E(k,\!q,\!\rho)}\right\}\non\\
&&\hskip 7cm+\frac{u|\eta|^2}{(f(0,\!\rho)-uq)^2}-uq \ .
\label{partial q}
\eea
By Lemma \ref{inf p2 rho} we have
\begin{equation*}
\frac{dp^{(2)}_\sLambda}{dq}(q,{\bar\rho}_\sLambda(q),\eta)
=\frac{\partial p^{(2)}_\sLambda}{\partial q}(q,{\bar\rho}_\sLambda(q),\eta)
+\frac{\partial p^{(2)}_\sLambda}{\partial \rho}(q,{\bar\rho}_\sLambda(q),\eta)
\frac{d{\bar\rho}_\sLambda(q)}{d q}
=\frac{\partial p^{(2)}_\sLambda}{\partial q}(q,{\bar\rho}_\sLambda(q),\eta)\,,
\end{equation*}
since $\partial p^{(2)}_{\sLambda}(q,{\bar\rho}_\sLambda(q),\eta)/{\partial \rho} = 0$. Therefore,
we can also write
\be
\label{identity=0}
\frac{dp^{(2)}_\sLambda}{dq}(q,{\bar\rho}_\sLambda(q),\eta)
=\frac{\partial p^{(2)}_\sLambda}{\partial q}(q,{\bar\rho}_\sLambda(q),\eta)
+ \frac{\partial p^{(2)}_\sLambda}{\partial \rho}(q,{\bar\rho}_\sLambda(q),\eta) \ .
\ee
Insertion of (\ref{partial rho}) and (\ref{partial q}) into the identity (\ref{identity=0}) gives
\begin{eqnarray}
\label{deriv-q for p-2 lambda}
\frac{dp^{(2)}_\sLambda}{d q}(q,{\bar\rho}_\sLambda(q),\eta) &=&
-\, \frac{1}{V}\sum_{k\in \Lambda^*}
\Big\{\ \frac{1}{\exp\{\beta E(k,\!q,\!{\bar\rho}_\sLambda(q))\}-
1}\ \frac{v f(k,\!{\bar\rho}_\sLambda(q))-u^2 q |\lambda(k)|^2}
{E(k,\!q,\!{\bar\rho}_\sLambda(q))}\non\\
&&\hskip 2cm+\half\(\frac{vf(k,\!{\bar\rho}_\sLambda(q))-
u^2 q |\lambda(k)|^2}{E(k,\!q,\!{\bar\rho}_\sLambda(q))}-v\)\Big\}\non \\
&&\hskip 4cm-\frac{\alpha|\eta|^2}{(f(0,\!{\bar\rho}_\sLambda(q))-uq)^2}+v{\bar\rho}_\sLambda(q)-uq \,.
\end{eqnarray}
Then, since $f(k,\!\rho)>u q |\lambda(k)|\geq u q |\lambda(k)|^2$,
$f(k,\!\rho)>E(k,\!q,\!{\bar\rho}_\sLambda(q))$ and
$\alpha>0 $, by (\ref{deriv-q for p-2 lambda}) we get the estimate
\be \label{estim-virt}
\frac{dp^{(2)}_\sLambda}{d q}(q,{\bar\rho}_\sLambda(q),\eta) \leq \, \frac{1}{2V}\sum_{k\in \Lambda^*}
\frac{u^2 q |\lambda(k)|^2}{E(k,\!q,\!{\bar\rho}_\sLambda(q))}
-\frac{\alpha|\eta|^2}{(f(0,\!{\bar\rho}_\sLambda(q))-uq)^2}+v{\bar\rho}_\sLambda(q)-uq \,.
\ee
Now we have
\bea
E^2(k,\!q,\!\rho)&=& (f(k,\!\rho)-u q |\lambda(k)|)(f(k,\!\rho)+u q |\lambda(k)|)\non\\
&=& (\eps(k)+\{f(0,\!\rho)-uq\}+uq\{1-|\lambda(k)|\})\non\\
&&\hskip 4cm \times(\eps(k)+\{f(0,\!\rho)-uq\}+uq\{1+|\lambda(k)|\})\non\\
&\geq& (f(0,\!\rho)-uq)uq \ . \label{estim-virt-1}
\eea
Therefore, by (\ref{lambda bound}), (\ref{M})  and (\ref{estim-virt}), (\ref{estim-virt-1}) we obtain
\be
\frac{dp^{(2)}_\sLambda}{dq}(q,{\bar\rho}_\sLambda(q),\eta)
<\frac{\mathfrak{C} \, {\mathfrak m}_\sLambda \, q^{1/2}u^{1/2}}{2V(f(0,\!{\bar\rho}_\sLambda(q))-
uq)^{1/2}}
-\frac{\alpha|\eta|^2}{(f(0,\!{\bar\rho}_\sLambda(q))-uq)^2}+
f(0,\!{\bar\rho}_\sLambda(q))-uq +\mu.
\label{upp bnd part q}
\ee
Let ${\ds \sigma_\sLambda(q):=\(f(0,\!{\bar\rho}_\sLambda(q))-uq\)(\max(1,q))^{1/3}}$. Then the
inequality (\ref{upp bnd part q}) gives
\be
\frac{dp^{(2)}_\sLambda}{dq}(q,{\bar\rho}_\sLambda(q),\eta)
<\frac{(\max(1,q))^{2/3}}{\sigma_\sLambda^{1/2}(q)}
\left\{\frac{\mathfrak{C} \, M \, u^{1/2}}{2}-\frac{\alpha|\eta|^2}{\sigma^{3/2}_\sLambda(q)} +
\sigma^{3/2}_\sLambda(q)\right\} +\mu.
\label{uppbnd part q 2}
\ee
Therefore, there exists $c_0(\eta)$ such that if $q\geq1$ and
$\sigma_\sLambda(q)<c_0(\eta)$, then
${\ds\frac{dp^{(2)}_\sLambda}{dq}(q,{\bar\rho}_\sLambda(q),\eta)<0}$ for all $\Lambda$.
Thus for all $\Lambda$ the
\textit{\textit{supremum}} of
$p^{(2)}_\sLambda(q,{\bar\rho}_\sLambda(q),\eta)$ over $q$ cannot be
attained in the domain defined by the condition $\sigma_\sLambda(q)< c_0(\eta)$.
\par
Now assume that $q\geq 1$ and $\sigma_\sLambda(q)\geq c_0(\eta)$.
Then, using again (\ref{estim-virt-1}),
we obtain from (\ref{partial q}) the estimate
\bea
\frac{\partial p^{(2)}_\sLambda}{\partial
q}(q,{\bar\rho}_\sLambda(q),\eta) &\leq& K
\left\{\frac{1}{(f(0,\!{\bar\rho}_\sLambda(q))-uq)}+
\frac{q^{1/2}}{(f(0,\!{\bar\rho}_\sLambda(q))-uq)^{1/2}}\right\}\non\\
&&\hskip 5cm +\frac{u|\eta|^2}{(f(0,\!{\bar\rho}_\sLambda(q))-uq)^2}-uq\non\\
&\leq& K \left\{\frac{q^{1/3}}{c_0(\eta)}+\frac{q^{2/3}}{c_0^{1/2}(\eta)}\right\}
+\frac{u|\eta|^2q^{2/3}}{c^2_0(\eta)}-uq \ .
\label{partial q b}
\eea
Since the right-hand side of (\ref{partial q b}) becomes negative for
large $q$, there is $q_0(\eta)<\infty$ such that the
\textit{\textit{supremum}} of $p^{(2)}_\sLambda(q,{\bar\rho}_\sLambda(q),\eta)$ with respect to
$q$  is attained in $q<q_0(\eta)$ for all $\Lambda$. Note that from
(\ref{partial q}) we see that if ${\bar q}_\sLambda$ is a maximizer
of $p^{(2)}_\sLambda(q,{\bar\rho}_\sLambda(q),\eta)$, then ${\bar
q}_\sLambda\neq 0$, and therefore combining this with the last
statement we can deduce that
\be
\frac{dp^{(2)}_\sLambda}{d q}({\bar
q}_\sLambda,{\bar\rho}_\sLambda({\bar q}_\sLambda),\eta)=0 \ .
\ee
Putting ${\bar c}_0(\eta)=c_0(\eta)/\{\max(1,q_0(\eta))\}^{1/3}$
finishes the proof.
\epr
\blem \label{L}
If $\eta\neq 0$, then
\be\label{lim-lemma4.4}
\lim_\Lambda \{p^{(2)}_\sLambda({\bar q}_\sLambda,{\bar \rho}_\sLambda({\bar q}_\sLambda),\eta)
-p^{(1)}_\sLambda({\bar q}_\sLambda e^{{\rm i}(\pi+2\psi)},\eta)\}=0 \ .
\ee
\elem
\begin{proof}
By Bogoliubov's inequality (\ref{J}) one gets
\bea
\label{Bog-ineq}
0&\leq&
p^{(2)}_\sLambda({\bar q}_\sLambda, {\bar \rho}_\sLambda({\bar
q}_\sLambda),\eta)- p^{(1)}_\sLambda({\bar q}_\sLambda e^{{\rm i}(\pi+2\psi)},\eta)\non\\
&=&{\tilde p}^{(2)}_\sLambda({\bar q}_\sLambda e^{{\rm i}(\pi+2\psi)},
{\bar \rho}_\sLambda({\bar q}_\sLambda),\eta)
- p^{(1)}_\sLambda({\bar q}_\sLambda e^{{\rm i}(\pi+2\psi)},\eta)\non \\
&\leq&
\frac{1}{2V^2}v\<(N_\sLambda-V {\bar \rho}_\sLambda({\bar
q}_\sLambda))^2\>_{H^{(2)}_\sLambda({\bar q}_\sLambda e^{{\rm i}(\pi+2\psi)},\, {\bar
\rho}_\sLambda({\bar q}_\sLambda),\eta)} \,.
\eea
Let $\delta N_\sLambda:=N_\sLambda -V{\bar \rho}_\sLambda({\bar q}_\sLambda)$ and
\be
\label{delta}
{\tilde\Delta}_\sLambda (\eta):=\<\delta
N^2_\sLambda\>_{H^{(2)}_\sLambda({\bar q}_\sLambda e^{{\rm i}(\pi+2\psi)},\, {\bar
\rho}_\sLambda({\bar q}_\sLambda),\eta)} \  .
\ee
Then (\ref{Bog-ineq}) implies
\begin{equation*}
0\leq
p^{(2)}_\sLambda({\bar q}_\sLambda, {\bar \rho}_\sLambda({\bar
q}_\sLambda),\eta)-p^{(1)}_\sLambda({\bar q}_\sLambda e^{{\rm i}(\pi+2\psi)},\eta)\leq
\frac{v}{2V^2}{\tilde\Delta}_\sLambda (\eta) \ .
\end{equation*}
We want to obtain an estimate for $ {\tilde\Delta}_\sLambda (\eta)$
in terms of $V$. To this end we introduce
\be \label{N-fluctuations}
{\tilde D}_\sLambda(\eta) =\(\delta N_\sLambda ,\, \delta
N_\sLambda\)_{H^{(2)}_\sLambda({\bar q}_\sLambda e^{{\rm
i}(\pi+2\psi)},\, {\bar \rho}_\sLambda({\bar q}_\sLambda),\eta)}
\ee
and calculate the derivatives
\bea
\frac{\partial p^{(2)}_\sLambda}{\partial
\mu}(q,\rho,\eta)&=&\frac{1}{V}\sum_{k\in \Lambda^*} \left \{\
\frac{1}{\exp(\beta E(k,\!q,\!\rho))-1}\
\frac{f(k,\!\rho)}{E(k,\!q,\!\rho)}
+\frac{1}{2}\(\frac{f(k,\!\rho)}{E(k,\!q,\!\rho)}-1\)\right\}\non \\
&&\hskip 7cm +\frac{v|\eta|^2}{(f(0,\!\rho)-uq)^2}\,,
\label{mu}
\eea
\be
\frac{\partial p^{(2)}_\sLambda}{\partial  \rho }(q,\rho,\eta)=
-v\(\frac{\partial p^0_\sLambda}{\partial \mu}
(q,\rho,\eta)-\rho\)\,,
\label{partial mu1}
\ee
\bea
&&\frac{\partial^2 p^{(2)}_\sLambda}{\partial \mu^2}(q,\rho,\eta)= \non \\
&&\frac{1}{V}\sum_{k\in \Lambda^*}
\left \{\ \frac{\beta\exp(\beta E(k,\!q,\!\rho))}{\(\exp(\beta E(k,\!q,\!\rho))-1\)^2}
\ \frac{f^2(k,\rho)}{E^2(k,\!q,\!\rho)} +\frac{1}{2}\
\frac{\exp(\beta E(k,\!q,\!\rho))+1}{\exp(\beta E(k,\!q,\!\rho))-1}\
\frac{u^2 q^2 |\lambda(k)|^2}{E^3(k,\!q,\!\rho)}\right\}\non \\
&&\hskip 7cm + \frac{2|\eta|^2}{(f(0,\!\rho)-uq)^3}\,.
\label{partial mu2} \eea From (\ref{partial mu2}), using
$e^x/(e^x-1)\leq 2(1+1/x) $ for $x\geq 0$ and
$f^2(k,\!\rho)=E(k,\!q,\!\rho)^2+u^2 q^2 |\lambda(k)|^2$, we get the
estimate \bea &&\frac{\partial^2 p^{(2)}_\sLambda}{\partial
\mu^2}(q,\rho,\eta) \leq \frac{2}{V}\sum_{k\in
\Lambda^*}\frac{1}{\(\exp(\beta E(k,\!q,\!\rho))-1\)} \(\beta
+\frac{1}{E(k,\!q,\!\rho)}\)
\non\\
&&\hskip 0cm +\frac{1}{V}\sum_{k\in \Lambda^*}\Bigg \{\frac{1}{\(\exp(\beta E(k,\!q,\!\rho))-1\)}\
\frac{2\beta E(k,\!q,\!\rho)+3}{E^3(k,\!q,\!\rho)}+
\frac{1}{2E^3(k,\!q,\!\rho)}\Bigg\}u^2 q^2 |\lambda(k)|^2\non \\
&&\hskip 9.5cm +\frac{2|\eta|^2}{(f(0,\!\rho)-uq)^3} \ .
\label{partial mu2 b}
\eea
The second sum in
(\ref{partial mu2 b}) is bounded from above by
\be
\frac{K_0}{V}\sum_{k\in \Lambda^*}\(\frac{1}{E^3(k,\!q,\!\rho)}+
\frac{1}{E^4(k,\!q,\!\rho)}\)u^2q^2|\lambda(k)|\leq
C\(\frac{1}{(f(0,\!\rho)-uq)^3}+\frac{1}{(f(0,\!\rho)-uq)^4}\)q^2 \,, \non
\ee
and the first sum (using $E^2(k,\!q,\!\rho)\geq \eps(k)(\eps(k)-\mu)$) by
\bea
&&\frac{K_{01}}{V}\left\{\sum_{{k\in \Lambda^*}\atop{\eps(k) \leq 1+4|\mu|/3}}
\(\frac{1}{E(k,\!q,\!\rho)}+\frac{1}{E^2(k,\!q,\!\rho)}\)
+\sum_{{k\in
\Lambda^*}\atop{\eps(k)>1+4|\mu|/3}}\frac{1}{\(\exp(\beta \eps(k)/2)-1\)}\right\}\non\\
&&\leq K_{02} \(\frac{1}{(f(0,\!\rho)-uq)}+\frac{1}{(f(0,\!\rho)-uq)^2} +1\)\,.\non
\eea
Consequently
\be
\frac{\partial^2 p^{(2)}_\sLambda}{\partial
\mu^2}({\bar q}_\sLambda,{\bar \rho}_\sLambda({\bar
q}_\sLambda),\eta) \leq
C_1\(\frac{1}{{\bar c}_0(\eta)}+\frac{1}{{\bar c}^2_0(\eta)}+\frac{q_0^2(\eta)}{{\bar c}^3_0(\eta)}
+\frac{q_0^2(\eta)}{{\bar c}^4_0(\eta)}+1\)+\frac{2|\eta|^2}{{\bar c}^3_0(\eta)} \ ,
\label{bound for partial mu2}
\ee
where ${\bar c}_0(\eta)$ and $q_0^2(\eta)$ are as in Lemma \ref{sup p2 q}.
\\
By Lemma \ref{sup p2 q} we have ${\ds\frac{\partial p^{(2)}_\sLambda}{\partial  \rho }({\bar
q}_\sLambda,{\bar\rho}_\sLambda({\bar q}_\sLambda),\eta)=0}$.
Then from (\ref{partial mu1}) one gets that
\begin{equation*}
{\bar\rho}_\sLambda({\bar q}_\sLambda)=
\frac{\partial p^{(2)}_\sLambda}{\partial  \mu } ({\bar
q}_\sLambda,{\bar\rho}_\sLambda({\bar q}_\sLambda),\eta)
=\frac{\partial {\tilde p}^{(2)}_\sLambda}{\partial  \mu }
({\bar q}_\sLambda e^{{\rm i}(\pi +2\psi)},{\bar\rho}_\sLambda({\bar q}_\sLambda),\eta)
=
\<\frac{N_\sLambda}{V}\>_{H^{(2)}_\sLambda({\bar q}_\sLambda,
{\bar \rho}_\sLambda({\bar q}_\sLambda),\eta)}\,,
\end{equation*}
and therefore by (\ref{N-fluctuations})
\begin{equation*}
\frac{{\tilde D}_\sLambda(\eta)}{V} =\frac{\partial^2 {\tilde
p}^{(2)}_\sLambda} {\partial \mu^2}({\bar q}_\sLambda e^{{\rm i}(\pi
+2\psi)},{\bar \rho}_\sLambda({\bar q}_\sLambda),\eta)
=\frac{\partial^2 p^{(2)}_\sLambda}{\partial \mu^2}({\bar
q}_\sLambda,{\bar \rho}_\sLambda({\bar q}_\sLambda),\eta) \,.
\end{equation*}
It then follows from (\ref{bound for partial mu2}) that
\be
\lim_{\Lambda}\frac{{\tilde D}_\sLambda(\eta)}{V^2}=0 \ .
\label{D(eta)}
\ee
Now Ginibre's inequality for (\ref{delta}) and (\ref{N-fluctuations}), cf. Section 3,
gives
\bea \label{Delta(eta)}
&&{\tilde \Delta}_\sLambda(\eta)\leq {\tilde D}_\sLambda(\eta)+   \\
\non && \half\beta^{1/2} \left\{{\tilde D}_\sLambda(\eta)\right\}^{1/2}
\left\{\<  [N_\sLambda,\
[H^{(2)}_\sLambda({\bar q}_\sLambda e^{{\rm i}(\pi +2\psi)},
{\bar \rho}_\sLambda({\bar q}_\sLambda),\eta),N_\sLambda]] \>_{H^{(2)}_\sLambda({\bar q}_\sLambda
e^{{\rm i}(\pi +2\psi)},
{\bar \rho}_\sLambda({\bar q}_\sLambda),\eta)}
\right\}^{1/2}.
\eea
Note that here
\bes
\<  [N_\sLambda,[H^{(2)}_\sLambda(q,\rho,\eta),N_\sLambda]] \>_{H^{(2)}_\sLambda(q,\rho,\eta)} =
2 u \< q^*Q^\*_\sLambda+ q Q^*_\sLambda \>_{H^{(2)}_\sLambda(q,\rho,\eta)}
+ {\sqrt V}\< \eta a^*_0 +\eta^* a^\*_0\>_{H^{(2)}_\sLambda(q,\rho,\eta)}.
\ees
By differentiating the pressure we find that
\bes
u \< q^*Q^\*_\sLambda+ q Q^*_\sLambda \>_{H^{(2)}_\sLambda(q,\rho,\eta)}
=2u|q|^2 V+\frac{2V}{u}\(q\frac{\partial {\tilde p}^{(2)}_\sLambda}{\partial q}(q,\rho,\eta)
+q^*\frac{\partial {\tilde p}^{(2)}_\sLambda}{\partial q^*}(q,\rho,\eta)\)\,,
\ees
so that if we define ${\hat q}:=|q|e^{{\rm i}(\pi +2\psi)}$, then we get
\bes
u \< {\hat q}^*Q^\*_\sLambda+ {\hat q} Q^*_\sLambda \>_{H^{(2)}_\sLambda({\hat q},\rho,\eta)}
=2u|q|^2 V+\frac{4V}{u}\(|q|\frac{\partial p^{(2)}_\sLambda}{\partial |q|}(|q|,\rho,\eta)\)\, .
\ees
An explicit calculation gives
\beas
\< \eta a^*_0 +\eta^* a^\*_0 \>_{H^{(2)}_\sLambda(q,\rho,\eta)}
&=& \sqrt{V}\(\eta\frac{\partial {\tilde p}^{(2)}_\sLambda}{\partial \eta}(q,\rho,\eta)
+\eta^*\frac{\partial {\tilde p}^{(2)}_\sLambda}{\partial \eta^*}(q,\rho,\eta) \)  \\
&=& 2|\eta|^2\sqrt{V}\left\{\frac{f(0,\!\rho)-u|q|\cos(\theta-2\psi)}{f^2(0,\!\rho)-u^2|q|^2}\right\}
\eeas
and so
\begin{equation}\label{BEC-1}
\< \eta a^*_0 +\eta^* a^\*_0 \>_{H^{(2)}_\sLambda({\hat q},\rho,\eta)}
= 2\sqrt{V}\left\{\frac{|\eta|^2}{f(0,\!\rho)-u|q|}\right\}.
\end{equation}
Therefore,  if ${\ds {\partial p^{(2)}_\sLambda}(|q|,\rho,\eta)/{\partial |q|}=0}$, then
\bes
\<[N_\sLambda,\
[H^{(2)}_\sLambda({\hat q},\rho,\eta),N_\sLambda]]\>_{H^{(2)}_\sLambda({\hat q},\rho,\eta)}
=2 V \,\(2u|q|^2+\frac{|\eta|^2}{(f(0,\!\rho)-u|q|)}\).
\ees
Thus
\begin{equation*}
\< [N_\sLambda,\ [H^{(2)}_\sLambda({\bar
q}_\sLambda e^{{\rm i}(\pi +2\psi)},{\bar \rho}_\sLambda({\bar q}_\sLambda),
\eta),N_\sLambda]] \>_{H^{(2)}_\sLambda({\bar
q}_\sLambda e^{{\rm i}(\pi +2\psi)},{\bar \rho}_\sLambda({\bar q}_\sLambda),\eta)} \leq
2V\(uq_0^2(\eta)+\frac{|\eta|^2}{{\bar c}_0(\eta)}\).
\end{equation*} 
From (\ref{D(eta)}), (\ref{Delta(eta)}) and the last estimate we
then see that
\begin{equation*}
\lim_{\Lambda}\frac{{\tilde
\Delta}_\sLambda(\eta)}{V^2}=0 \,,
\end{equation*}
completing the proof.
\end{proof}

Now we prove that the order of the thermodynamic limit and taking
the \textit{\textit{infimum}} and \textit{\textit{supremum}} in (\ref{lim-lemma4.4}) can be
reversed.
\par
{\bf Proof of Theorem \ref{interchange Theor 2nd}\,:}
\par
We know from Lemma \ref{sup p2 q} that there is $q_0(\eta)<\infty$, independent of $\Lambda$, such
that for large $\Lambda$, the maximizer ${\bar q}_\sLambda\in[0,q_0(\eta)]$.
Then it follows from Lemma \ref{inf p2 rho}
that
$\delta_0(\eta):=\inf_{q\in[0,\,q_0(\eta)]}v{\tilde\rho}_1(q,\!\eta)-(\mu + u{\bar q})_+ >0$ and
${\tilde\rho}_{02}(\eta):=\sup_{q\in[0,\,q_0(\eta)]}{\tilde\rho}_2(q,\!\eta)<\infty$.
Thus ${\bar \rho}_\sLambda({\bar q})$
is in $[0,{\tilde\rho}_{0 2}(\eta)]$
and  $v{\bar \rho}_\sLambda({\bar q})-(\mu + u{\bar q})_+ > \delta_0(\eta)$. Let
$G_\eta\subset \RR_+^2$ be the  compact set
\be
G_\eta:=\{(q,\rho)\
|\ 0\leq q \leq q_0(\eta),\ [(\mu + u q)_+ +\delta_0(\eta)]/v\leq
\rho \leq {\tilde\rho}_{0 2}(\eta)\} \ .\non
\ee
Then $({\bar q}_\sLambda,{\bar \rho}_\sLambda({\bar q}_\sLambda))\in G_\eta$.
Therefore, there is a sequence $\Lambda_n$ such that
$({\bar q}_{\sLambda_n},{\bar \rho}_{\sLambda_n}({\bar q}_{\sLambda_n})$
converges to some point $({\bar q},{\bar \rho})$ in $G_\eta$.
\par
The derivatives of $p_\sLambda^{(2)}(q,\rho,\eta)$ are uniformly
bounded on $G_\eta$ and therefore as $\Lambda\uparrow \RR^\nu$, $p_\sLambda^{(2)}(q,\rho,\eta)$
converges uniformly to $p^{(2)}(q,\rho,\eta)$ on $G_\eta$. Thus
\be
\label{lim-p-2}
\lim_\Lambda p_\sLambda(\eta)= \lim_{n\to\infty}
p_{\sLambda_n}^{(2)}({\bar q}_{\sLambda_n}, {\bar
\rho}_{\sLambda_n}({\bar q}_{\sLambda_n}),\eta) =p^{(2)}({\bar
q},{\bar \rho},\eta) \ .
\ee
By repeating the arguments of Lemmas \ref{inf p2 rho} and \ref{sup p2 q} and by replacing
(for $V \rightarrow \infty$) the sums over $k$ by integrals,
we see that if ${\bar{\bar q}}$ maximizer of $\inf_{\rho\,:\,\sigma(q,\rho)\geq 0}p^{(2)}(q,\rho,\eta)$
with respect to $q$, then $ 0\leq {\bar{\bar q}} \leq q_0(\eta)$ and if
${\bar{\bar \rho}}({\bar{\bar q}})$
is a minimizer of $p^{(2)}({\bar{\bar q}},\rho,\eta)$, then
$({\bar{\bar q}},{\bar{\bar \rho}}({\bar{\bar q}}))$ is in $G_\eta$.
Thus
\be
\sup_{q\geq
0}\inf_{\rho\,:\,\sigma(q,\rho)\geq 0}p^{(2)}(q,\rho,\eta)=\sup_{q\in
[0,q_0(\eta)]}\inf_{\{\rho : (q,\rho)\in
G_\eta\}}p^{(2)}(q,\rho,\eta) \,.\non
\ee
Since
\be
p_{\sLambda_n}^{(2)}({\bar
q}_{\sLambda_n}, {\bar \rho}_{\sLambda_n}({\bar
q}_{\sLambda_n}),\eta) \leq p_{\sLambda_n}^{(2)}({\bar
q}_{\sLambda_n},\rho,\eta) \non
\ee
for $\rho$ such that $({\bar q}_{\sLambda_n},\rho)\in G_\eta$, we get also that
\be p^{(2)}({\bar
q},{\bar \rho},\eta) \leq p^{(2)}({\bar q},\rho,\eta)\,,\non
\ee
for $\rho$ such that $({\bar q},\rho)\in G_\eta$. That is
\be
\label{inf-p-2}
p^{(2)}({\bar
q},{\bar \rho},\eta)= \inf_{\{\rho : ({\bar q},\rho)\in
G_\eta\}}p^{(2)}({\bar q},\rho,\eta) \ .
\ee
Similarly, for all $q\geq 0$ we have
\be
\label{ineq-p-2}
p_{\sLambda_n}^{(2)}({\bar q}_{\sLambda_n}, {\bar
\rho}_{\sLambda_n}({\bar q}_{\sLambda_n}),\eta) \geq
p_{\sLambda_n}^{(2)}(q,{\bar \rho}_{\sLambda_n}(q),\eta) \ .
\ee
If $0\leq q \leq  q_0(\eta)$, then $(q,{\bar \rho}_{\sLambda_n}(q))\in
G_\eta$ and therefore ${\bar \rho}_{\sLambda_n}(q)$ has a
convergent subsequence ${\bar \rho}_{\sLambda_{n_r}}(q)$
converging to some ${\hat \rho}$, where $(q,{\hat \rho})\in G_\eta$.
Taking the limit in (\ref{ineq-p-2}) we obtain
\be
p^{(2)}({\bar q},{\bar \rho},\eta)\geq
p^{(2)}(q,{\hat \rho},\eta) \geq \inf_{\{\rho : (q,\rho)\in
G_\eta\}}p^{(2)}(q,\rho,\eta). \non
\ee
Thus, by (\ref{inf-p-2})
\be
p^{(2)}({\bar q},{\bar
\rho},\eta) = \inf_{\{\rho : ({\bar q},\rho)\in
G_\eta\}}p^{(2)}({\bar q},\rho,\eta)\geq \inf_{\{\rho : (q,\rho)\in
G_\eta\}}p^{(2)}(q,\rho,\eta) \non
\ee
for all $q\in[0,q_0(\eta)]$.
Therefore
\be
\label{press-2-lim}
p^{(2)}({\bar q},{\bar \rho},\eta) =
\sup_{q\in[0,q_0(\eta)]}\inf_{\{\rho : (q,\rho)\in
G_\eta\}}p^{(2)}(q,\rho,\eta) =\sup_{q\geq 0}\inf_{\rho\,:\,\sigma(q,\rho)\geq 0}p^{(2)}(q,\rho,\eta).
\ee
Combining the relation (\ref{press-2-lim}) with (\ref{lim-p-2}) we prove the theorem and obtain an
explicit formula for the limiting value of the pressure. \qquad \hfill $\square$

\begin{remark}\label{press-2-sup-inf>0}
By the definition of ${\bar \rho}_{\sLambda}(q)$, ${\bar q}_{\sLambda}$
(Lemma \ref{inf p2 rho}, \ref{sup p2 q}) and by (\ref{lim-p-2}) we also get that for
$\eta \neq 0$
\begin{eqnarray}
&&\lim_{n\to\infty}
p_{\sLambda_n}^{(2)}({\bar q}_{\sLambda_n}, {\bar
\rho}_{\sLambda_n}({\bar q}_{\sLambda_n}),\eta)=\lim_{n\to\infty}
\sup_{q \geq 0} \inf_{\rho \geq 0} p_{\sLambda_n}^{(2)}(q,
\rho, \eta) = \sup_{q \geq 0}\inf_{\rho \geq 0} p^{(2)}(q, \rho, \eta) \non \\
&& = p^{(2)}({\bar q},{\bar \rho},\eta)\,,\label{lim-p-2-sup-inf}
\end{eqnarray}
where (cf. (\ref{diag p-real}))
\begin{eqnarray}\label{diag p-real-lim}
p^{(2)}(q,\rho,\eta)&=& - \int_{\RR^\nu} \frac{d^\nu k}{(2 \pi)^\nu}\left
\{\frac{1}{\beta } \ln \left[1-\exp(-\beta E(k,\!q,\!\rho))\right]
+ \frac{1}{2}\(E(k,\!q,\!\rho) -f(k,\!\rho)\)\right\}\non \\
&+& \frac{|\eta|^2}{f(0,\!\rho)-u q} - \frac{1}{2}uq^2 +\frac{1}{2}v\rho^2 \,,
\end{eqnarray}
and ${\bar q},{\bar \rho}$ satisfy the equations
\begin{equation}\label{q,rho-eq}
\frac{\partial p^{(2)}}{\partial  \rho }(q,\rho,\eta) = 0 \,\,\,\,\,\,,\,\,\,\,\,\,
\frac{\partial p^{(2)}}{\partial  q }(q,\rho,\eta) = 0 \,.
\end{equation}
\end{remark}
We now show that the zero-mode $\eta$-source term can be \textit{switched off}.
\blem \label{off source}
Thermodynamic limit of the pressure is equal to
\begin{equation*}
p:=\lim_{\Lambda \uparrow \RR^\nu}p_\sLambda=\lim_{\eta \to 0 }
\lim_{\Lambda \uparrow \RR^\nu}p_\sLambda(\eta) \ .
\end{equation*}
\elem
\begin{proof}
By Bogoliubov's  convexity inequality (\ref{J}) one gets \be
-\frac{|\eta|}{\sqrt V}|\<a^\*_0+a^*_0)\>_{H_\sLambda}|\leq
p_\sLambda-p_\sLambda(\eta)\leq \frac{|\eta|}{\sqrt
V}|\<a^\*_0+a^*_0)\>_{H_\sLambda(\eta)}|\,, \non \ee that implies
\be 0\leq \left|p_\sLambda-p_\sLambda(\eta)\right|\leq
\frac{2|\eta|}{\sqrt V}|\<a^*_0\>_{H_\sLambda(\eta)}|\leq
\frac{2|\eta|}{\sqrt V}\<a^*_0a^\*_0\>^\half_{H_\sLambda(\eta)} \leq
\frac{2|\eta|}{\sqrt V}\<N_\sLambda\>^\half_{H_\sLambda(\eta)}.
\label{Q} \ee From Lemma \ref{App1} and (\ref{H}) we see that for
$|\eta|\leq 1$,
\begin{equation*}
\< \frac{N_\sLambda}{V}\>_{H_\sLambda(\eta)}\leq K_1 \ ,
\end{equation*}
where $K_1$ is independent of $\eta$. Thus the right-hand side of (\ref{Q}) tends to zero as $\eta$
tends to zero.
\end{proof}
Finally we prove that the order of the limit $\eta\to 0$ and taking
the \textit{\textit{infimum}} and \textit{\textit{supremum}} in (\ref{press-2-lim}) can be
\textit{reversed}.
\begin{lemma} \label{Var Princ}
\begin{equation}\label{revers-lim}
\lim_{\eta\to 0}\sup_{q\geq 0}\inf_{\rho\,:\,\sigma(q,\rho)\geq 0}p^{(2)}(q,\rho,\eta)
= \sup_{q\geq 0}\inf_{\rho\,:\,\sigma(q,\rho)\geq 0}p^{(2)}(q,\rho) \ ,
\end{equation}
where $p^{(2)}(q,\rho):= p^{(2)}(q,\rho,0)$ is defined in (\ref{lim-press-2nd Approx-eta=0}).
\end{lemma}
\begin{proof}
Let ${\bar \rho}_\eta(q)$ be such that
\be
\inf_{\rho\,:\,\sigma(q,\rho)\geq 0}p^{(2)}(q,\rho,\eta)=p^{(2)}(q,{\bar \rho}_\eta(q),\eta) \ , \non
\ee
and ${\bar q}_\eta$ be such that
\be \sup_{q\geq 0}p^{(2)}(q,{\bar \rho}_\eta(q),\eta)= p^{(2)}({\bar
q}_\eta,{\bar \rho}_\eta({\bar q}_\eta),\eta) \ . \non
\ee
Let
\be
G_0:=\{(q,\rho)\ |\ q \geq0 \,, \sigma(q,\rho)\geq 0\} \ . \non
\ee
By arguments similar to the above (see proof of Theorem \ref{interchange Theor 2nd})
we can show that these exist and that
$({\bar q}_\eta,{\bar \rho}_\eta({\bar q}_\eta))\in G_0$. We shall
need the following derivative of (\ref{diag p-real-lim}):
\bea
&&\frac{\partial
p^{(2)}}{\partial  \rho }(q,\rho,\eta) =-v\int_{\RR^\nu} \frac{d^\nu k}{(2 \pi)^\nu}\
\left \{\ \frac{1}{\exp(\beta E(k,\!q,\!\rho))-1}\
\frac{f(k,\!\rho)}{E(k,\!q,\!\rho)}
+\frac{1}{2}\(\frac{f(k,\!\rho)}{E(k,\!q,\!\rho)}-1\)\right\}\non \\
&&\hskip 3cm- \frac{v|\eta|^2}{(f(0,\!\rho)-uq)^2}+
v\rho \, .
\label{dir1}
\eea
Moreover, in the same way as in (\ref{partial q}), (\ref{deriv-q for p-2 lambda}) we also obtain:
\bea
&& \frac{dp^{(2)}}{d q}(q,{\bar\rho}_\eta(q),\eta)
=\non \\
&& \hskip 1cm u^2 q\int_{\RR^\nu}\hskip -0.2cm \frac{d^\nu k}{(2 \pi)^\nu}\,
|\lambda(k)|^2\left \{\ \frac{1}{\exp(\beta E(k,\!q,\!{\bar\rho}_\eta(q)))-1}
\ \frac{1}{E(k,\!q,\!{\bar\rho}_\eta(q))}
+  \frac{1}{2E(k,\!q,\!{\bar\rho}_\eta(q))}\right\}\non\\
&&\hskip 7cm+ \frac{u|\eta|^2}{(f(0,\!{\bar\rho}_\eta(q))-uq)^2}-uq \,,
\label{dir2}
\eea
and for any number $t$
\bea
\frac{dp^{(2)}}{d q}(q,{\bar\rho}_\eta(q),\eta)&=& -\int_{\RR^\nu}
\hskip -0.2cm \frac{d^\nu k}{(2 \pi)^\nu}\,
\Bigg \{\frac{1}{\exp(\beta E(k,\!q,\!{\bar\rho}_\eta(q)))-1}\
\frac{t\,v f(k,\!{\bar\rho}_\eta(q))-u^2 q|\lambda(k)|^2}{E(k,\!q,\!{\bar\rho}_\eta(q))}\non\\
&&\hskip 2cm +\half\(\frac{t\,vf(k,\!{\bar\rho}_\eta(q))-{u^2 q|\lambda(k)|^2}}
{E(k,\!q,\!{\bar\rho}_\eta(q))}-t\,v\)
\hskip -0.1cm\Bigg\}\non \\
&&\hskip 4cm- \frac{\alpha|\eta|^2}{(f(0,\!{\bar\rho}_\eta(q))-uq)^2}
+t\,v{\bar\rho}_\eta(q)-uq \,.
\label{dir3}
\eea
As in (\ref{partial q b}), from (\ref{dir2}) we get the estimate
\begin{eqnarray}
\frac{d p^{(2)}}{d q}(q,{\bar\rho}_\eta(q),\eta)
&\leq& K \left\{\frac{1}{(f(0,\!{\bar\rho}_\eta(q))-uq)}
+ \frac{q^{1/2}}{(f(0,\!{\bar\rho}_\eta(q))-uq)^{1/2}}\right\} \non \\
&+& \frac{u|\eta|^2}{(f(0,\!{\bar\rho}_\eta(q))-uq)^2}-uq \,.\label{estim-der-p2-q}
\end{eqnarray}
Therefore, if $f(0,\!{\bar \rho}_\eta({\bar q}_\eta))-u{\bar
q}_\eta\geq 1$, then by the definition of ${\bar q}_\eta$ and by (\ref{estim-der-p2-q}) we obtain
\be
0 = \frac{d p^{(2)}}{d q}({\bar q}_\eta,{\bar
\rho}_\eta({\bar q}_\eta),\eta) \leq \frac{K_1(1+
{\bar q}_\eta^{1/2})}{(f(0,\!{\bar \rho}_\eta({\bar q}_\eta))-u{\bar
q}_\eta)^{1/2}} -u{\bar q}_\eta \ .\non
\ee
Since the right-hand side of the last inequality must be non-negative, then
\be
f(0,\!{\bar \rho}_\eta({\bar q}_\eta))-u{\bar q}_\eta \leq
\frac{K_1^2(1+{\bar q}_\eta^\half)^2}{u^2{\bar q}^2_\eta} \ . \non
\ee
Similarly, if $f(0,\!{\bar \rho}_\eta({\bar q}_\eta))-u{\bar q}_\eta\leq 1$, then
\bes
\frac{d p^{(2)}}{d q}({\bar q}_\eta,{\bar \rho}_\eta({\bar
q}_\eta),\eta) \leq \frac{K_2(1+{\bar q}_\eta^{1/2})}
{(f(0,\!{\bar \rho}_\eta({\bar q}_\eta))-u{\bar q}_\eta)^2} -u{\bar q}_\eta \,.
\ees
The right-hand side of the last inequality must be positive and thus
\bes f(0,\!{\bar \rho}_\eta({\bar q}_\eta))-u{\bar q}_\eta \leq
\frac{K_2^{1/2}(1+{\bar q}_\eta^{1/2})^{1/2}}{u^{1/2}{\bar q}^{1/2}_\eta} \ .
\ees
Therefore, either
\bes
1\leq f(0,\!{\bar \rho}_\eta({\bar q}_\eta))-u{\bar q}_\eta \leq
\frac{K_1^2(1+{\bar q}_\eta^{1/2})^2}{u^2{\bar q}^2_\eta} \ \ \ \text{or}\ \ \
0\leq f(0,\!{\bar \rho}_\eta({\bar q}_\eta))-u{\bar q}_\eta \leq \min\(1,
\frac{K_2^{1/2}(1+{\bar q}_\eta^{1/2})^{1/2}}{u^{1/2}{\bar q}^{1/2}_\eta}\) \ .
\ees
Thus the only way that $({\bar q}_\eta,{\bar
\rho}_\eta({\bar q}_\eta))$ can escape to infinity as $\eta\to 0$ is, if
either ${\bar \rho}_\eta({\bar q}_\eta)\to \infty$ and ${\bar q}_\eta\to
0$, or if ${\bar \rho}_\eta({\bar q}_\eta)\to \infty$, ${\bar q}_\eta\to
\infty$ and $f(0,\!{\bar \rho}_\eta({\bar q}_\eta))-u{\bar
q}_\eta\to 0$. Now, if $\rho\to \infty$ and $q\to 0$, the right-hand
side of (\ref{dir1}) tends to $+\infty$. Therefore the case
${\bar \rho}_\eta({\bar q}_\eta)\to \infty$ and ${\bar q}_\eta\to 0$, is not possible.

Suppose now that ${\bar \rho}_\eta({\bar q}_\eta)\to \infty$, ${\bar q}_\eta\to
\infty$ and $f(0,\!{\bar \rho}_\eta({\bar q}_\eta))-u{\bar
q}_\eta\to 0$. From (\ref{dir3}) with $t=u/v$ we get
\bes
 0 = \frac{dp^{(2)}}{d q}({\bar q}_\eta,{\bar\rho}_\eta({\bar q}_\eta),\eta)
<\frac{\|\lambda\|u}{2} +u{\bar\rho}_\eta({\bar q}_\eta)) -u{\bar q}_\eta
=\frac{\|\lambda\|u}{2} +\frac{u}{v}\(f(0,\!{\bar\rho}_\eta({\bar q}_\eta))-u{\bar q}_\eta +
\mu-\alpha{\bar q}_\eta\).
\ees
This contradicts our supposition and therefore ${\bar\rho}_\eta({\bar q}_\eta)$ and
${\bar q}_\eta$ must remain \textit{finite}.
\par
As in (\ref{upp bnd part q}) and (\ref{uppbnd part q 2}), from (\ref{dir3}) with $t=1$, we get
\beas
 0 = \frac{dp^{(2)}}{d q}({\bar q}_\eta,{\bar\rho}_\eta({\bar q}_\eta),\eta)
&<&\frac{1}{(f(0,\!{\bar\rho}_\eta({\bar q}_\eta))-u{\bar q}_\eta)^{1/2}}
\(\frac{\|\lambda\|u^{1/2}{\bar q}_\eta^{1/2}}{2}-
\frac{\alpha|\eta|^2}{(f(0,\!{\bar\rho}_\eta({\bar q}_\eta))-u{\bar q}_\eta)^{3/2}}\)\\
&&\hskip 4cm +f(0,\!{\bar\rho}_\eta({\bar q}_\eta))-u{\bar q}_\eta +\mu.
\eeas
Therefore, since the right-hand side must be positive,  the term
\begin{equation*}
\frac{|\eta|^2}{(f(0,\!{\bar\rho}_\eta({\bar q}_\eta))-u{\bar q}_\eta)^{3/2}}
\end{equation*}
must remain bounded when  $f(0,\!{\bar \rho}_\eta({\bar q}_\eta))-
u{\bar q}_\eta\to 0$.
\par
Summarizing we see that $({\bar q}_\eta,{\bar\rho}_\eta({\bar q}_\eta))$ must remain in a
\textit{bounded
subset} of $G_0$ and
\be
\label{lim-eta-to-zero}
\lim_{\eta\rightarrow 0} \,\,\, \frac{|\eta|^2}{(f(0,\!{\bar\rho}_\eta({\bar q}_\eta)-u{\bar q}_\eta)}
= 0 \,.
\ee
Since $({\bar q}_\eta,{\bar\rho}_\eta({\bar q}_\eta))$
remains in a bounded subset of $G_0$, there exists a sequence $\eta_n\to 0$ such that
$({\bar q}_{\eta_n},{\bar \rho}_{\eta_n}({\bar q}_{\eta_n}))$
converges to $({\bar q},{\bar \rho})\in {\bar G}_0$, where ${\bar G}_0$ is the closure of $G_0$.
Now $p^{(2)}(q,\rho)$ is continuous on ${\bar G}_0$.
Thus by (\ref{lim-eta-to-zero}) we obtain
\bea
\lim_\Lambda p_\sLambda &=&
\lim_{n\to\infty} p^{(2)}({\bar q}_{\eta_n},{\bar \rho}_{\eta_n}({\bar q}_{\eta_n}),\eta_n)\non\\
&=&
\lim_{n\to\infty} p^{(2)}({\bar q}_{\eta_n},{\bar \rho}_{\eta_n}({\bar q}_{\eta_n}))+
\lim_{n\to\infty}\frac{|\eta|^2}{(f(0,\!{\bar\rho}_{\eta_n}({\bar q}_{\eta_n})-
u{\bar q}_{\eta_n})}\non\\
&=& p^{(2)}({\bar q},{\bar \rho}) \ . \non \eea Now for $\rho$ such
that $({\bar q},\rho)\in G_0$, for large $n$ we have $({\bar
q}_{\eta_n},\rho)\in G_0$. Therefore, for large $n$ we get \be
p^{(2)}({\bar q}_{\eta_n}, {\bar \rho}_{\eta_n}({\bar
q}_{\eta_n}),\eta_n) \leq p^{(2)}({\bar q}_{\eta_n},\rho,\eta_n)
\non \ee and letting $n\to\infty$, we obtain for $\rho$ such that
$({\bar q},\rho)\in G_0$, the estimate \be p^{(2)}({\bar q},{\bar
\rho})\leq p^{(2)}({\bar q},\rho) \ . \non \ee That is \be
p^{(2)}({\bar q},{\bar \rho})= \inf_{\{\rho : ({\bar q},\rho)\in
G_0\}}p^{(2)}({\bar q},\rho).\non \ee Similarly,  for all $q\geq 0$
we have \be p_{\eta_n}^{(2)}({\bar q}_{\eta_n}, {\bar
\rho}_{\eta_n}({\bar q}_{\eta_n}),\eta_n) \geq
p_{\eta_n}^{(2)}(q,{\bar \rho}_{\eta_n}(q),\eta_n) \ . \non \ee From
(\ref{dir1}), we see that for each  $q\geq 0$, both ${\bar
\rho}_\eta(q)$  and $|\eta|^2/(f(0,\!{\bar\rho}_\eta(q)-uq)^2$
remain bounded as $\eta \to 0$. Let $\left\{{\bar
\rho}_{\eta_{n_r}}(q)\right\}_{n_r \geq 1}$ be a convergent
subsequence of $\left\{{\bar \rho}_{\eta_n}(q)\right\}_{n\geq 1}$
converging to ${\hat \rho}$ say,  where $(q,{\hat \rho})\in {\bar
G}_0$. By letting $r\to\infty$ we then have \be p^{(2)}({\bar
q},{\bar \rho})\geq p^{(2)}(q,{\hat \rho}) \geq \inf_{\{\rho :
(q,\rho)\in G_0\}}p^{(2)}(q,\rho) \ . \non \ee Therefore \be
p^{(2)}({\bar q},{\bar \rho}) = \inf_{\{\rho : ({\bar q},\rho)\in
G_0\}}p^{(2)}({\bar q},\rho)\geq \inf_{\{\rho : (q,\rho)\in
G_0\}}p^{(2)}(q,\rho), \non \ee for all $q\geq 0$, and thus we get
the relation \be p^{(2)}({\bar q},{\bar \rho}) = \sup_{q\geq
0}\inf_{\{\rho : (q,\rho)\in G_0\}}p^{(2)}(q,\rho) = \sup_{q\geq
0}\inf_{\rho\,:\,\sigma(q,\rho)\geq 0}p^{(2)}(q,\rho) \non \ee
proving the theorem. \end{proof}
\par
Combining Theorem \ref{interchange Theor 2nd}, Lemma \ref{off source} and Lemma \ref{Var Princ}
we get the \textit{first} part of our main result, Theorem \ref{Main Theorem}, (\ref{lim-Var Princ}).\\
The second part we shall consider in the next section.

\section{Discussion}\label{Discussion}

Let us put in Hamiltonian (\ref{PBH-sources}) the source equal $\nu =0$ and suppose that $\eta
\neq 0$. Then the corresponding \textit{Euler-Lagrange
equations}, obtained by the condition that the derivatives (\ref{dir1}) and
(\ref{dir2}) are equal to zero, take the form
\begin{eqnarray}
\rho =\half
\int_{\RR^\nu} \frac{d^\nu k}{(2 \pi)^\nu}\ \left
\{\frac{f(k,\!\rho)}{{E}(k,\!q,\!\rho)}\ \coth \half \beta
{E}(k,\!q,\!\rho)-1\right\}  &+&
\frac{|\eta|^2}{(f(0,\!\rho)- uq)^2} \ , \label{EL1} \\
q =\frac{u
\, q}{2} \int_{\RR^\nu} \frac{d^\nu k}{(2 \pi)^\nu}\
\frac{|\lambda(k)|^2}{{E}(k,\!q,\!\rho)} \ \coth \half \beta
{E}(k,\!q,\!\rho)\  &+& \frac{|\eta|^2}{(f(0,\!{\rho})- uq)^2} \ . {\label{EL2}}
\end{eqnarray}
We shall now discuss some of the consequences of these equation in relation to the
existence of Bose-Einstein condensation (BEC) in the model (\ref{PBH-sources}).
\par
(a) The solution $({\bar \rho}_\eta (\beta, \mu), {\bar q}_\eta
(\beta, \mu))$ of the equations (\ref{EL1}), (\ref{EL2}) always exist and is a smooth
function of $\beta, \mu$ and $\eta$, for $\eta \neq 0$. Moreover, we can
identify it with the Gibbs expectations of the corresponding observables.
Since the pressure $p_{\Lambda}(\nu=0,\eta)$ is a \textit{convex}
function of $\mu$ and of $u$, then by the \textit{Griffiths lemma}, see e.g.
\cite{ZagBru}, the corresponding derivatives converges in the
thermodynamic limit to derivatives of the limiting pressure
(\ref{lim-pres-eta}). Differentiating (\ref{lim-pres-eta}) with
respect to $\mu$ and $u$ and comparing these derivatives with the
solutions of (\ref{EL1}) and (\ref{EL2}), we get
\bes
\lim_{\Lambda}\<\frac{N_\sLambda}{V}\>_{H_\sLambda(\nu=0,
\eta)}={\bar \rho}_\eta (\beta, \mu), \ \ \ \ \lim
_{\Lambda}\<\frac{Q^*_\sLambda
Q^\*_\sLambda}{V^2}\>_{H_\sLambda(\nu=0, \eta)}= {\bar q}^2_\eta
(\beta, \mu).
\ees
(b) Similarly we can show that the zero-mode BEC for $\eta\neq 0$ is given by
\be
\label{zero-mode-cond-BEC-equal}
\rho_0(\eta):=\lim_{\Lambda} \<\frac{{a}^*_0{a}^\*_0 }{{V}}\>_{{H}_\sLambda(0, \eta)}
=
\frac{|\eta|^2}{(f(0,\!{\bar \rho}_\eta)-u{\bar q}_{\eta})^2}\,.
\ee
To obtain this result let us make a \textit{global} gauge transformation
$U_\varphi = e^{i\varphi N_\Lambda}$ of the Hamiltonian
$H_\sLambda(\mu, \nu=0,\eta)= H_\sLambda(\nu=0,\eta)- \mu N_\sLambda
$, see (\ref{PBH-sources}),  with $\varphi = \arg \eta$. Then :
\begin{equation*}
\tilde{H}_\sLambda(\mu, 0, \eta)= U_\varphi H_\sLambda(\mu, 0,\eta) U_{\varphi}^* =
\tilde{H}_\sLambda - \mu N_\sLambda - {\sqrt V} |\eta| \(\tilde{a}^*_0 + \tilde{a}^\*_0\) \ .
\end{equation*}
From
\begin{equation*}
0= \langle[\tilde{H}_\sLambda(\mu, 0, \eta), N_\sLambda ]\rangle_{\tilde{H}_\sLambda(\mu,
0, \eta)}
= {\sqrt V} |\eta| \langle\tilde{a}^*_0 - \tilde{a}^\*_0 \rangle_{\tilde{H}_\sLambda(\mu, 0, \eta)}
\end{equation*}
and
\begin{equation*}
0 \leq \langle[N_\sLambda ,
[\tilde{H}_\sLambda(\mu, 0, \eta), N_\sLambda ]]\rangle_{\tilde{H}_\sLambda(\mu,
0, \eta)}
= {\sqrt V} |\eta| \langle\tilde{a}^*_0 + \tilde{a}^\*_0 \rangle_{\tilde{H}_\sLambda(\mu, 0, \eta)}
\end{equation*}
we obtain
\begin{equation}\label{zero-mode}
\langle \tilde{a}^*_0 \rangle_{\tilde{H}_\sLambda(\mu, 0, \eta)}=
\langle \tilde{a}_0 \rangle_{\tilde{H}_\sLambda(\mu, 0, \eta)} \geq 0 .
\end{equation}
Let $\delta A_0 := (\tilde{a}^*_0 + \tilde{a}^\*_0) -
\langle\tilde{a}^*_0 + \tilde{a}^\*_0 \rangle_{\tilde{H}_\sLambda(\mu, 0, \eta)}$. Then
\begin{equation}\label{press-conv}
\frac{\partial^2 p_{\sLambda}(\eta)}{\partial |\eta|^2} = \(\delta A_{0}^* ,\
\delta A_{0}\)_{\tilde{H}_\sLambda(\mu, 0, \eta)} \geq 0 ,
\end{equation}
where $(\cdot\ ,\ \cdot)_{\tilde{H}_\sLambda(\mu, 0, \eta)}$
denotes the Bogoliubov-Duhamel inner
product with respect to the Hamiltonian $\tilde{H}_\sLambda(\mu, \nu =0, \eta)$.
Hence, the convexity (\ref{press-conv}) and convergence of the pressure $p_{\sLambda}(\eta)$ (see
Theorem \ref{interchange Theor 2nd} and Remark \ref{press-2-sup-inf>0}) imply by the
Griffiths lemma the convergence of the
first derivatives to the derivative of the limiting pressure :
\begin{equation}\label{BEC-2}
\lim_{\Lambda} \frac{\partial p_{\sLambda}(\eta)}{\partial |\eta|}
= \lim_{\Lambda} \frac{1}{\sqrt{V}}\langle\tilde{a}^*_0 +
\tilde{a}^\*_0 \rangle_{\tilde{H}_\sLambda(\mu, 0, \eta)} =
\frac{{2|\eta|}}{f(0,\!{\bar \rho}_\eta)-u {\bar q}_\eta} \,,
\end{equation}
see (\ref{lim-pres-eta}), (\ref{press-real-q-eta-lim}) and (\ref{BEC-1}). Therefore,
by (\ref{zero-mode}), (\ref{BEC-2}), and returning back to original zero-mode operators,
we obtain
\begin{equation}\label{BEC-3}
\lim_{\Lambda}\<\frac{a^*_0}{\sqrt{V}}\>_{H_\sLambda(0, \eta)}=
\frac{{\eta}^*}{f(0,\!{\bar \rho}_\eta)-u {\bar q}_\eta} \ \ \ , \ \ \
\lim_{\Lambda}\<\frac{a_0}{\sqrt{V}}\>_{H_\sLambda(0, \eta)}=
\frac{{\eta}}{f(0,\!{\bar \rho}_\eta)-u {\bar q}_\eta} \ .
\end{equation}
So, by (\ref{BEC-3}) we conclude that the $\eta$ - source in Hamiltonian (\ref{PBH-sources})
\textit{breaks} the zero-mode gauge invariance creating a zero-mode macroscopic
occupation with the particle density  estimated from below by the Cauchy-Schwarz inequality:
\begin{eqnarray}
\label{zero-mode-cond-estim}
\lim _{\Lambda}\<\frac{a^*_0 a^\*_0 }{V}\>_{H_\sLambda(0, \eta)} &\geq&
\lim _{\Lambda}\<\frac{a^*_0}{\sqrt{V}}\>_{H_\sLambda(0, \eta)} \
\< \frac{a^\*_0 }{\sqrt{V}}\>_{H_\sLambda(0, \eta)} \\
&=& \frac{|\eta|^2}{(f(0,\!{\bar \rho}_\eta)-u {\bar q}_\eta)^2}\,.\nonumber
\end{eqnarray}
To prove that in fact there is an \textit{equality} in
(\ref{zero-mode-cond-estim}),
we consider $p_{\sLambda}(\eta,s)$ the pressure with $\epsilon(0)$ replaced by $\epsilon(0)-s$
with $s$ positive and  again use its convexity with respect to $s$. Then Griffiths lemma and that fact that
$f(0,\!{\bar \rho}_\eta)-u{\bar q}_{\eta} > 0$, as soon as $\eta \neq 0$, imply, see
(\ref{diag p-real}) and (\ref{lim-p-2}):
\bes \label{zero-mode-cond-BEC}
\lim_{\Lambda} \<\frac{{a}^*_0{a}^\*_0 }{{V}}\>_{{H}_\sLambda(0, \eta)}
=
\lim_{\Lambda} \ \left( \frac{\partial p_{\sLambda}(\eta,s)}{\partial s}\right)_{s=+0}
\leq
\left( \frac{\partial p(\eta,s)}{\partial s}\right)_{s=+0}
=
\frac{|\eta|^2}{(f(0,\!{\bar \rho}_\eta)-u{\bar q}_{\eta})^2}\,.
\ees
Here we have used the fact that the $s$-dependence of $p(\eta,s)$ is only through the last term in
(\ref{press-real-q-eta-lim}).
\par
(c) In the limit $\eta \rightarrow 0$ equations (\ref{EL1}) and
(\ref{EL2}) coincide with equations (3.7) and (3.8) or (3.10) and
(3.11) in \cite{PZ-Pair}.  There the amount of the \textit{generalized
condensate} density  is denoted there by $m_0(\beta, \mu)$.
By inspection this coincides with the limit of $\rho_{0}(\eta)$ in (\ref{zero-mode-cond-BEC-equal}) as $\eta\to 0$:
\begin{equation*}
m_0(\beta, \mu)=
\lim_{\eta \rightarrow 0}\rho_{0}(\eta) \ .
\end{equation*}
In \cite{PZ-Pair} we found that for $m_0 $ to be non-zero, $\mu$ must be greater than a certain
\textit{critical value} of chemical potential $\mu_c(\beta, u,v)$ .
For $u=0$, this critical chemical potential coincides with the one
for the Mean-Field boson gas (\ref{MF-Hamil}), namely
$\mu_c(\beta, u=0,v) = v\rho_c(\beta)$,  where $\rho_c(\beta)$ is the \textit{critical
density} for the Perfect Bose-gas, see e.g. \cite{PZ-Mean}.

(d) It was shown in \cite{PZ-Pair} that the phase diagram is
quite complicated. Subject to these Euler-Lagrange equations the
expressions  for the pressure given in \cite{PZ-Pair} equation (2.11) and at
the top of page 438, are the same as $p^{(2)}(q,\rho)$ in
(\ref{lim-press-2nd Approx-eta=0}). (We warn the reader that in
these equations for the pressure in \cite{PZ-Pair} there is a
misprint and a term is missing.) There we were able to solve the
problem only for some values of $u$ and $v$, see Fig. 2  in
\cite{PZ-Pair}. For example (\ref{EL2}) shows that for $u >0$
(\textit{attraction} in the BCS part of the PBH (\ref{hamiltonian}))
the existence of the generalized Bose condensate $m_0 \neq 0$ causes
an \textit{abnormal boson pairing}:
\begin{equation}\label{abn-pair}
\lim_{\eta \rightarrow 0}\lim _{\Lambda}
\frac{1}{2}\<Q_\sLambda^* + Q_\sLambda\>_{H_\sLambda(0, \eta)}=
\lim_{\eta \rightarrow 0} {\bar q}_\eta(\beta,\mu) \neq 0 \,.
\end{equation}
This is because, for $u >0$, equation (\ref{EL2}) cannot have the trivial solution
${\bar q}_{\eta} = 0$ when the \textit{generalized} condensate
\begin{equation}\label{GBCond}
m_0(\beta, \mu) = \lim_{\eta \rightarrow 0} \frac{|\eta|^2}{(f(0,\!{\bar
\rho}_\eta)-u{\bar q}_{\eta})^2} \neq 0 \,.
\end{equation}
Note that on the other hand the equations (\ref{EL1}) and
(\ref{EL2}) allow the possibility that $m_0 = 0$ without $\lim_{\eta
\rightarrow 0}{\bar q}_\eta = 0$. This \lq\lq two-stage"
condensation is possible only when $u >0$ and it is similar to that
discussed in \cite{PZ-Pair}.
\par
(e) As in \cite{PZ-Pair} we interpret the spectrum (\ref{E}) of the effective Hamiltonian
\begin{equation}\label{spec-eta-to-zero}
\varepsilon_{\rm excit}(k):= \lim_{\eta \rightarrow0}E(k,\!\overline{q}_\eta,\!\overline{\rho}_\eta) \,,
\end{equation}
as the \textit{spectrum of excitations} for the PBH (\ref{hamiltonian}).
 Our analysis of the Euler-Lagrange equations (\ref{EL1}), (\ref{EL2}) (as well as
(\ref{EL1-w}), (\ref{EL2-w} below) shows that there \textit{no gap} in this spectrum
as soon as there is the Bose condensation (\ref{GBCond}):
\begin{equation}\label{gap-zero}
\lim_{k \rightarrow 0}\varepsilon_{\rm excit}(k)=\lim_{k \rightarrow 0}\lim_{\eta \rightarrow0}(\epsilon(k)
- \mu + \overline{\rho}_\eta -
|u \overline{q}_\eta \lambda^{*}(k)|) = 0 .
\end{equation}
This conclusion is again in agreement with \cite{PZ-Pair}.
\par
(f) The case of \textit{repulsion} ($u \leq 0$) in the BCS part of the PBH (\ref{hamiltonian})
is quite different. In this case the pressure coincides with the \textit{mean-field} one ($u=0$)
and we always have for the \textit{boson paring}: $\lim_{\eta\rightarrow 0} {\bar q}_\eta (\beta, \mu) = 0 $.
The first property was derived in great generality in \cite{PZ-Pair}. To make a contact
with the variational principle proved in this paper, let us change
notation and replace $u$ by $- w$,  with  $w \geq 0$. The Euler-Lagrange equations,
(\ref{EL1})and (\ref{EL2}), become
\begin{eqnarray}
\rho =\half \int_{\RR^\nu} \frac{d^\nu k}{(2
\pi)^\nu}\ \left \{\frac{f(k,\!\rho)}{{E}(k,\!q,\!\rho)}\ \coth
\half \beta {E}(k,\!q,\!\rho)-1\right\} &+&
\frac{|\eta|^2}{(f(0,\, \rho)+ wq)^2} \ , \label{EL1-w}\\
q =\frac{(- w) q}{2} \int_{\RR^\nu} \frac{d^\nu k}{(2 \pi)^\nu}\
\frac{{|\lambda(k)|^2}}{{E}(k,\!q,\!\rho)} \ \coth \half \beta
{E}(k,\!q,\!\rho) &+& \frac{|\eta|^2}{(f(0,\, \rho)+ wq)^2} \ .\label{EL2-w}
\end{eqnarray}
Since the solutions ${\bar \rho}_{\eta}(\beta, \mu)$, ${\bar q}_{\eta}(\beta, \mu)$
of equations (\ref{EL1-w}), (\ref{EL2-w})
must satisfy the condition $\sigma({\bar q}_{\eta}, {\bar
\rho}_{\eta})\geq 0$,  one gets by (\ref{inf sigma}) the estimate
\begin{equation}\label{est-below}
f(0, \, {\bar \rho}_{\eta})+ w {\bar q}_{\eta} \geq 2 w {\bar q}_{\eta} \ .
\end{equation}
Note that the first term in the right-hand side of (\ref{EL2-w}) is negative. Therefore, by (\ref{est-below})
we obtain
\begin{equation*}
{\bar q}_{\eta}(\beta, \mu)<\frac{|\eta|^2}{(f(0,\, \rho)+ w{\bar q}_{\eta}(\beta, \mu))^2}
<\frac{|\eta|^2}{(2w{\bar q}_{\eta}(\beta, \mu))^2}\ \ \
\textrm{or}\ \ \  {\bar q}_{\eta}(\beta, \mu)<\frac{|\eta|^{2/3}}{(2w)^{2/3}} \ .
\end{equation*}
This implies that in the limit $\eta\rightarrow 0$ the equation
(\ref{EL2-w}) may have  only a \textit{trivial} solution:
\begin{equation}\label{q-zero}
\lim_{\eta \rightarrow 0} {\bar q}_{\eta}(\beta, \mu) =0  \ ,
\end{equation}
and
\begin{equation}\label{zero-mode-zero}
\lim_{\eta \rightarrow 0}\frac{|\eta|^2}{(f(0,\, {\bar\rho}_{\eta})
+ w {\bar q}_{\eta})^2} = 0 \ .
\end{equation}
Let $\rho_c (\beta)$ be the critical density for the \textit{Perfect} Bose Gas:
$w=v=0$, see (\ref{N-Q-oper}) or (\ref{MF-Hamil}),
\begin{equation*}
\rho_c (\beta):=\int_{\RR^\nu} \frac{d^\nu k}{(2 \pi)^\nu}\
\frac{1}{e^{\beta \eps(k)}-1}.
\end{equation*}
For $\mu\leq v\rho_c(\beta)$, limits (\ref{q-zero}) and (\ref{zero-mode-zero}) imply that as $\eta \to 0$
the solution of equation (\ref{EL1-w})
tends to $\hat{\rho}(\beta,\mu)$ the solution of the corresponding equation for the Mean-Field model (\ref{MF-Hamil}):
\bes
\rho = \int_{\RR^\nu} \frac{d^\nu k}{(2\pi)^\nu}\ \frac{1}{e^{\beta(\eps(k)- \mu + v \rho)}-1} \ \  ,
\ees
and the pressure
\begin{equation*}
p^{w}(\beta, \mu):= \lim_{\eta \rightarrow 0}\inf_{\rho\,:\,\sigma(q,\rho)\geq
0}\inf_{q\geq 0}p^{(2)}(q,\rho,\eta)=
\inf_{\rho\,:\,\sigma(0,\rho)\geq 0}p^{(2)}(0,\rho,0) = p^{(2)}(0,\hat{\rho}(\beta,\mu))
\end{equation*}
coincides with the mean-field pressure, see (\ref{lim-press-2nd Approx-eta=0}) and \cite{PZ-Mean}.
On the other hand, if $\rho >\mu/v$, then from (\ref{EL1-w}) we obtain for any $\varepsilon>0$ and
$\eta$ is sufficiently small
\begin{equation*}\label{}
\frac{\mu}{v}< {\bar \rho}_{\eta}(\beta, \mu)=\rho_c(\beta) +\varepsilon \ ,
\end{equation*}
giving a contradiction for $\mu>v\rho_c(\beta)$.
This means that in this case equations (\ref{EL1-w}) and (\ref{EL2-w}) are \textit{inconsistent} and the
minimum point must lie on the \textit{boundary} of the allowed range on the $\rho$-$q$ plane.
This boundary
consists of the two lines $q=0$ and $ \rho=(\mu+wq)/v$. Minimizing the pressure on the first
line is equivalent to solving the variational problem in the mean-field case. This was done
in \cite{PZ-Mean} where one sees that the minimum is attained at a point which tends to
$ \rho=\mu/v$ as $\eta\to 0$.
On the other boundary $\rho=(\mu+wq)/v$ similar calculations show that the
minimizer also tends to $(\rho=\mu/v,q=0)$. Thus the pressure again coincides with the
with the mean-field pressure.\\
This proves the \textit{second} part of our main result for repulsive BCS interaction in the PBH,
Theorem \ref{Main Theorem}, (\ref{lim-Var Princ-u negative}).
\par
We end with the following remark concerning BEC in the PBH model.
Though the pressure of the model with the PB Hamiltonian for $w>0$
coincides with the one for $w=0$, it is an open question wether these models
\textit{coincide} completely.  As has been shown in
\cite{MichVer}-\cite{BruZag2} a similar type of \textit{diagonal} quadratic repulsion is able to
change the type of Bose condensation, from condensation in the \textit{zero mode} (type I) to
 \textit{generalized} van den Berg-Lewis-Pul\'{e} condensation \cite{BLP} out of the zero mode
\textit{without altering} the pressure. Therefore, the analysis of the Bose condensate structure in the
PBH model requires a more detailed study of the corresponding
quantum Gibbs states. This is beyond the scope of the present paper.


\textbf{Acknowledgements}\\
J.V.P. wishes to thank the Centre de Physique Th\'eorique, Luminy-Marseille and
V.A.Z. the
School of Mathematical Sciences,
University College Dublin and the Dublin Institute for Advanced Studies
for their warm hospitality and financial support.
%
%
\section*{Appendix A: Commutators}\label{Appendix A}
\renewcommand{\theequation}{\Alph{section}.\arabic{equation}}
\setcounter{section}{1}
\setcounter{equation}{0}
\setcounter{theorem}{0}
By (\ref{N-Q-oper}) and  (\ref{PBH-sources}) we have
\bea
[H_\sLambda(\nu, \eta)-\mu N_\sLambda, Q^\*_\sLambda] &=& (- 2) \
\sum_{k\in \Lambda^*}(\eps(k)-\mu)\lambda(k)A_k
+ \frac{2u}{V} \ \sum_{k\in \Lambda^*}|\lambda(k)|^2 \left(N_{k}+
\frac{1}{2}\right) \ Q^\*_\sLambda \non\\
&-& \frac{v}{V}(N_\sLambda Q^\*_\sLambda \ + \ Q^\*_\sLambda N_\sLambda)
+ 4 \nu \ \sum_{k\in \Lambda^*}|\lambda(k)|^2 \left(N_{k}+
\frac{1}{2}\right) - 2{\sqrt V}\eta a^\*_0 \,,\non
\eea
and
\bea
\label{double comm}
[Q^*_\sLambda,[H_\sLambda(\nu, \eta)-\mu N_\sLambda, Q^\*_\sLambda]] &=&
8 \, \sum_{k\in \Lambda^*}(\eps(k)-\mu) |\lambda(k)|^2 \left(N_{k} +
\frac{1}{2}\right)\non\\
&-&\frac{4 u}{V}\left\{\sum_{k\in \Lambda^*}|\lambda(k)|^2 \lambda^{*}(k)A^{*}_k \ Q^\*_\sLambda
+ 2 \left[\sum_{k\in \Lambda^*} |\lambda(k)|^2 \left(N_{k} +
\frac{1}{2}\right)\right]^2\right\} \non\\
&+&\frac{4 v}{V}\left\{Q^*_\sLambda Q^\*_\sLambda +
2 \, \sum_{k\in \Lambda^*}|\lambda(k)|^2 \left( N_{k}+\frac{1}{2} \right) (N_\sLambda +1)\right\}\non \\
&-&8 \nu \ \sum_{k\in \Lambda^*}|\lambda(k)|^2 \lambda^{*}(k)A^{*}_k +
4 \sqrt{V} \eta \lambda(0) a^{*}_0 \, .
\eea
Using (\ref{M2}) and (\ref{M3}) we see that the first term in (\ref{double comm}) is bounded by
\begin{equation*}
8(\mathfrak{c}_\sLambda +|\mu|)\la N_\sLambda \ra + 4\mathfrak{n}_\sLambda +
4|\mu|\mathfrak{m}_\sLambda  \,,
\end{equation*}
where  $\la\, \cdot\, \ra := \la\, \cdot\, \ra_{H_\sLambda(\nu, \eta)}$.
Recall that Lemma \ref{Bounds on Q*Q} gives
\bes
Q^*_\sLambda Q^\*_\sLambda \leq N^2_\sLambda + M V N_\sLambda
\ees
and as in (\ref{ineq-A-N}) we get $A^\*_kA^*_k \leq N^\*_k N^\*_{-k}+3(N^\*_k +N^\*_{-k})+2$.
Using these we obtain
\beas
\sum_{k\in \Lambda^*}|\lambda(k)|^3|\la A^*_k \ Q^\*_\sLambda\ra| &\leq &
\sum_{k\in \Lambda^*}|\lambda^\*(k)|\la A^\*_kA^*_k\ra^{1/2}\la \ Q^*_\sLambda Q^\*_\sLambda\ra^{1/2}\\
&\leq & \la N^2_\sLambda + M V N_\sLambda \ra^{1/2}
\(\sum_{k\in \Lambda^*}|\lambda(k)|\)^{1/2}\(\sum_{k\in \Lambda^*}|\lambda(k)|\la A^\*_kA^*_k\ra\)^{1/2}\\
&\leq & \la N^2_\sLambda + M V N_\sLambda \ra^{1/2}
\mathfrak{m}_\sLambda^{1/2}\(\sum_{k\in \Lambda^*}|\lambda(k)|\la N^\*_k N^\*_{-k}+
3(N^\*_k +N^\*_{-k})+2\ra\)^{1/2}\\
&\leq & \la N^2_\sLambda + M V N_\sLambda \ra^{1/2}
\mathfrak{m}_\sLambda^{1/2}\(\la N^2_\sLambda+6N_\sLambda+2\mathfrak{m}_\sLambda\ra\)^{1/2},
\eeas
and independently we have
\bes
\sum_{k\in \Lambda^*} |\lambda(k)|^2 \left(N_{k} +\frac{1}{2}\right)
\leq
\sum_{k\in \Lambda^*} |\lambda(k)| \left(N_{k} +\frac{1}{2}\right)
\leq
N_\sLambda +\frac{\mathfrak{m}_\sLambda}{2}
\ees
and
\bes
\sum_{k\in \Lambda^*}|\lambda(k)|^2 \( N_{k}+\frac{1}{2} \) \(N_\sLambda +1\)
\leq
\sum_{k\in \Lambda^*}|\lambda(k)| \( N_{k}+\frac{1}{2} \) \(N_\sLambda +1\)
\leq
\(N_\sLambda +\frac{\mathfrak{m}_\sLambda}{2}\)\(N_\sLambda +1\),
\ees
which gives estimates for the second and the third terms in (\ref{double comm}).
We now bound the penultimate term in (\ref{double comm}).
\beas
\sum_{k\in \Lambda^*}|\lambda(k)|^3|\< A^*_k  \>|
&\leq &
\sum_{k\in \Lambda^*}|\lambda(k)||\< A^*_k  \>|\leq
\sum_{k\in \Lambda^*}\< N_{-k} \>^{1/2} \< N_k +1\>^{1/2} |\lambda(k)|^{1/2}\non\\
&\leq &
\(\sum_{k\in \Lambda^*}\< N_{-k} \>\)^{1/2}
\(\sum_{k\in \Lambda^*}|\lambda(k)|\< N_k +1\>\)^{1/2} \leq
\< N_\sLambda \>^{1/2} \(\< N_\sLambda \> +\mathfrak{m}_\sLambda\)^{1/2}.
\eeas
Finally for the last term we have
\be
|\<a^*_0 \> |\leq \< N_0 \>^{1/2} \leq \< N_\sLambda\>^{1/2} \,.\non
\ee
Putting these bounds together we get
\bea
\< [Q^*_\sLambda,[H_\sLambda(\nu, \eta)-\mu N_\sLambda, Q^\*_\sLambda]]\> &\leq&
8(\mathfrak{c}_\sLambda +|\mu|)\la N_\sLambda \ra + 4\mathfrak{n}_\sLambda +
4|\mu|\mathfrak{m}_\sLambda\non\\
&&+\frac{4u}{V}\la N^2_\sLambda + M V N_\sLambda \ra^{1/2}
\mathfrak{m}_\sLambda^{1/2}\(\la N^2_\sLambda+6N_\sLambda+2\mathfrak{m}_\sLambda\ra\)^{1/2}\non\\
&& + \frac{8u}{V}\(N_\sLambda +\frac{\mathfrak{m}_\sLambda}{2}\)^2
+\frac{4v}{V}\(N^2_\sLambda + M V N_\sLambda\) \non \\
&&+\frac{8v}{V}\(N_\sLambda +\frac{\mathfrak{m}_\sLambda}{2}\)\(N_\sLambda +1\)\non\\
&&+8 \< N_\sLambda \>^{1/2} \(\< N_\sLambda \>
+\mathfrak{m}_\sLambda\)^{1/2}+32{\sqrt V} \< N_\sLambda\>^{1/2}\,.
\non \eea for $|\nu|\leq 1$ and $|\eta|\leq 1$. From Lemma
\ref{App1} and (\ref{H}) we see that for $|\nu|\leq 1$ and
$|\eta|\leq 1$, \be \<\frac{N_\sLambda }{V}\>_{H_\sLambda(\nu,
\eta)}\leq K_1 \ \ \ \ {\rm and} \ \ \ \ \< \frac{N^2_\sLambda
}{V^2}\>_{H_\sLambda(\nu, \eta)}\leq K_2 , \ee where $K_1$ and $K_2$
are independent of $\nu , \eta$. Thus \bea
\<[Q^*_\sLambda,[H_\sLambda(\nu, \eta)-\mu N_\sLambda,
Q^\*_\sLambda]]\>_{H_\sLambda(\nu, \eta)} \leq C\, V^{3/2} \eea for
some number $C$.
\section*{Appendix B: Bounds}\label{Appendix B}
\setcounter{section}{2}
\setcounter{equation}{0}
\renewcommand{\theequation}{\Alph{section}.\arabic{equation}}
\renewcommand{\thelemma}{\Alph{section}.\arabic{lemma}}
\begin{lemma}
\label{App1}
If a Hamiltonian $ H_\sLambda $ satisfies the condition
\be
H_\sLambda\geq T_\sLambda+\frac{1}{2V}\gamma N^2_\sLambda -\delta N_\sLambda-\sigma V
\label{G}
\ee
with $\gamma>0$ then there exist constants $K_1$ and $K_2$,
depending only on $\gamma$, $\delta$, $\sigma$ and $\mu$ but not on $\Lambda$, such that
\be
\< \frac{N_\sLambda }{V}\>_{H_\sLambda}\leq K_1
\ee
and
\be
\< \frac{N^2_\sLambda }{V^2}\>_{H_\sLambda}\leq K_2 \,.
\ee
\end{lemma}
\begin{proof}
Let $p_\sLambda(\mu)$ be the pressure for $ H_\sLambda $, then
$$
\< \frac{N_\sLambda }{V}\>_{H_\sLambda}\leq p_\sLambda(\mu+1)
-p_\sLambda(\mu)\leq p_\sLambda(\mu+1)\leq K_1,
$$
where $K_1$ is independent of $\Lambda$ by (\ref{G}). Also for $\lambda \in [0,\gamma) $ let
$$
H_\sLambda(\lambda):=H_\sLambda-\frac{1}{2V}\lambda N^2_\sLambda \,,$$
and let $p_\sLambda(\mu,\lambda)$ be the corresponding pressure.
Then $$
\< \frac{N^2_\sLambda }{V^2}\>_{H_\sLambda}
\leq \frac{2}{\gamma}\{p_\sLambda(\mu, \gamma/2)-p_\sLambda(\mu)\}\leq \frac{2}
{\gamma}p_\sLambda(\mu, \gamma/2)\leq K_2 \,,
$$
where $K_2$ is independent of $\Lambda$, again by (\ref{G}).
\end{proof}
Note that by Theorem \ref{super} the Hamiltonians (\ref{hamiltonian}) and (\ref{PBH-sources})
verify the condition (\ref{G}), see estimate (\ref{H}).
%

\end{document}